\documentclass[aip,jcp,preprint,superscriptaddress,noshowkeys]{revtex4-1}
\usepackage{amssymb,physics,xcolor}
\usepackage{tikz}
\usepackage{float}
\usepackage{notes2bib}
\usepackage[margin=2.7cm]{geometry}
\usetikzlibrary{shapes.geometric, arrows}

\usepackage{hyperref}
\hypersetup{
    colorlinks,
    linkcolor={red!50!black},
    citecolor={red!70!black},
    urlcolor={red!80!black}
}

\tikzstyle{start} = [rectangle, rounded corners, minimum width=3cm, minimum height=1cm,text centered, draw=black, fill=red!30]
\tikzstyle{stop} = [rectangle, rounded corners, minimum width=3cm, minimum height=1cm,text centered, draw=black, fill=red!30]
\tikzstyle{point} = [rectangle, minimum width=0.000001cm, minimum height=0.0000001cm,text centered, draw=black, fill=white!30]
\tikzstyle{io} = [rectangle, rounded corners, minimum width=3cm, minimum height=1cm, text centered, draw=black, fill=blue!30]
\tikzstyle{process} = [rectangle, minimum width=3cm, minimum height=1cm, text centered, draw=black, fill=orange!30]
\tikzstyle{decision} = [rectangle, minimum width=3cm, minimum height=1cm, text centered, draw=black, fill=green!30]
\tikzstyle{arrow} = [thick,->,>=stealth]

\definecolor{auburn}{rgb}{0.43, 0.21, 0.1}

\newcommand{\rev}[1]{\textcolor{black}{#1}}

\newcommand{\anh}[2]{{\hat{#1}^{\vphantom{\dagger}}}_{#2}}
\newcommand{\cre}[2]{{\hat{#1}^{\dagger}}_{#2}}

\newcommand{\br}{\boldsymbol{r}}
\newcommand{\bc}{\boldsymbol{c}}

\newcommand{\bt}{\boldsymbol{t}}
\newcommand{\bz}{\boldsymbol{z}}
\newcommand{\bn}{\boldsymbol{n}}
\newcommand{\bkappa}{\boldsymbol{\kappa}}

\newcommand{\bF}{\boldsymbol{F}}
\newcommand{\bO}{\boldsymbol{0}}

\newcommand{\bG}{\boldsymbol{G}}
\newcommand{\bg}{\boldsymbol{g}}
\newcommand{\hH}{\hat{H}}
\newcommand{\hh}{\hat{h}}
\newcommand{\hZ}{\hat{Z}}
\newcommand{\stH}{\bar{H}}
\newcommand{\hS}{\hat{S}}
\newcommand{\hT}{\hat{T}}
\newcommand{\hTheta}{\hat{\Theta}}
\newcommand{\hkappa}{\hat{\kappa}}
\newcommand{\tG}{\tilde{G}}
\newcommand{\cK}{\mathcal{K}}
\newcommand{\hcK}{\hat{\cK}}
\newcommand{\ii}{\mathrm{i}}

\usepackage[version=4]{mhchem} % Formula subscripts using \ce{}
\usepackage{appendix}
\usepackage{caption}
\usepackage{subcaption}

\begin{document}

\title{Time-Reversal Symmetry in RDMFT and pCCD with Complex-Valued Orbitals}

\author{Mauricio Rodr\'iguez-Mayorga}
\email{marm3.14@gmail.com}
\affiliation
{Grenoble Alpes University, CNRS, Grenoble INP, Institut Néel, 25 rue des Martyrs, 38042 Grenoble, France}
\affiliation
{Department of Chemistry and Pharmaceutical Sciences, Vrije Universiteit, De Boelelaan 1108, 1081 HZ Amsterdam, The
Netherlands}
\author{Pierre-Fran\c{c}ois Loos}
\affiliation
{Laboratoire de Chimie et Physique Quantiques (UMR 5626), Université de Toulouse, CNRS, UPS, France}
\author{\\Fabien Bruneval}
\affiliation
{Universit\'e Paris-Saclay, CEA, Service de recherche en Corrosion et Comportement des
Mat\'eriaux, SRMP, 91191 Gif-sur-Yvette, France}
\author{Lucas Visscher}
\affiliation
{Department of Chemistry and Pharmaceutical Sciences, Vrije Universiteit, De Boelelaan 1108, 1081 HZ Amsterdam, The
Netherlands}

\begin{abstract}
Reduced density matrix functional theory (RDMFT) and coupled cluster theory restricted to paired double excitations (pCCD) are emerging as efficient methodologies for accounting for the so-called non-dynamic electronic correlation effects. Up to now, molecular calculations have been performed with real-valued orbitals. However, before extending the applicability of these methodologies to extended systems, where Bloch states are employed, the subtleties of working with complex-valued orbitals and the consequences of imposing time-reversal symmetry must be carefully addressed. In this work, we describe the theoretical and practical implications of adopting time-reversal symmetry in RDMFT and pCCD when allowing for complex-valued orbital coefficients. The theoretical considerations primarily affect the optimization algorithms, while the practical implications raise fundamental questions about the stability of solutions. Specifically, we find that complex solutions lower the energy when non-dynamic electronic correlation effects are pronounced. We present numerical examples to illustrate and discuss these instabilities and possible problems introduced by $N$-representability violations. 
\bigskip
\begin{center}
	\boxed{\includegraphics[scale=0.5, clip, trim=0.0cm 10.0cm 0.0cm 6.0cm]{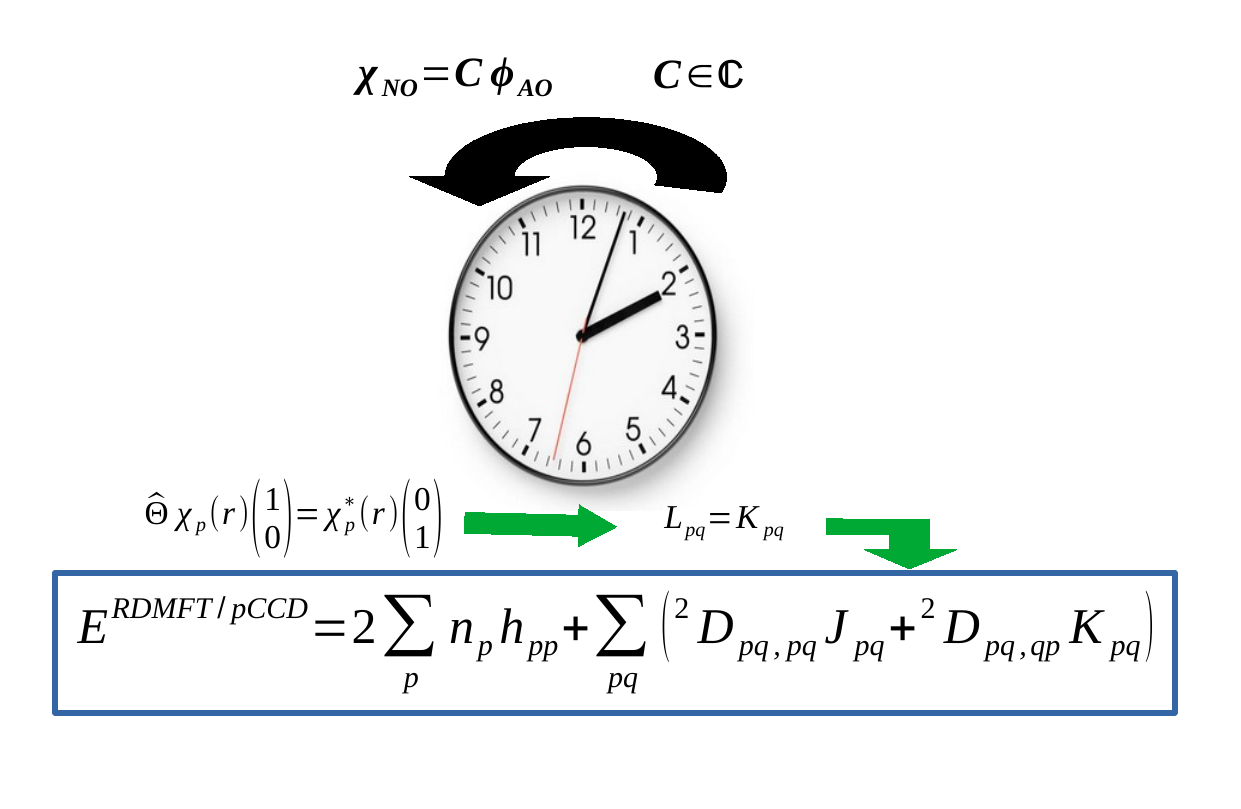}}
\end{center}
\bigskip
%\begin{figure}
%    \centering
%    \includegraphics[scale=0.7, clip, trim=0.0cm 10.0cm 0.0cm 6.0cm]{figures/toc.pdf}
%\end{figure}
\end{abstract}

\maketitle

\section{Introduction}
In quantum chemistry, accurately describing the so-called electronic correlation effects\cite{lowdin:55pr} remains an open problem.  For practical purposes, it has been convenient to classify and measure these effects as dynamic and non-dynamic, where the former can be interpreted as small corrections on top of the Hartree-Fock (HF) reference determinant and the latter refers to major changes in the electronic wave function caused by (near-)degeneracies in the single-particle states.~\cite{cioslowski:91pra,lee:89ijqc,lee:03cpl,ramos-cordoba:16pccp,ramos-cordoba:17jctc,via2019singling,boguslawski:12jpcl} Accounting for the so-called non-dynamic electronic correlation effects in quantum chemistry has been routinely tackled using multiconfigurational self-consistent field methodologies, \cite{helgaker:00book,malmqvist1989casscf} such as complete-active-space self-consistent field (CASSCF),\cite{roos1980complete,roos1980completeijqc,roos2005multiconfigurational} complete-active-space configuration interaction (CASCI), or density-matrix renormalization group (DMRG).~\cite{white1992density,white1993density,chan2011density,wouters2014density,szalay2015tensor,baiardi2020density} However, their applicability is limited due to the exponential growth of their computational cost with respect to the system size. 

Alternative methodologies like reduced density matrix functional theory \cite{gilbert:75prb,valone1991density} (RDMFT) and coupled-cluster theory restricted to paired double excitations \cite{limacher2013new} including its orbital-optimized version~\cite{boguslawski2014efficient,henderson2014seniority} (labeled as pCCD in this work) are recently gaining practitioners in the electronic structure community. \cite{piris:13ijqc,piris2018dynamic,piris:21gnof,mitxelena2020efficient1,mitxelena2020efficient2,mitxelena2024assessing,rodriguez2022relativistic,henderson2014seniority,boguslawski2016targeting,brzek2019benchmarking,boguslawski2021open,boguslawski2021orbital,boguslawski2024benchmarking,kossoski2023seniority,kossoski2021excited,ravi2023excited,marie2021variational,tecmer2022geminal,jahani2023relationship,ahmadkhani2024linear}${}^,$\bibnote{Notice that for the pCCD method, different orbital optimization schemes have been proposed in the literature~\cite{boguslawski2014projected,boguslawski2014nonvariational}; however, the one used in this work is taken from Ref.~\onlinecite{henderson2014seniority}.} Within these methodologies, the diagonalization of large matrices is replaced by the optimization of occupation numbers or amplitudes, which drastically reduces the computational cost. Furthermore, these methodologies are cost-effective approaches to deal with the so-called non-dynamic electronic correlation effects \cite{piris:13ijqc,mitxelena2020efficient1,mitxelena2020efficient2,henderson2014seniority,boguslawski:12jpcl,boguslawski2024benchmarking} because the optimization procedure introduces fractional occupation numbers that adjust to the degeneracies present in the system under investigation. For this reason, the most simple RDMFT approximations, the M\"uller \cite{muller:93jpc} and power \cite{sharma:08pow,cioslowski2000:power}  functionals, have already been employed to study strongly correlated materials such as nickel oxides, \cite{sharma:08pow,sharma2013spectral} where these methods describe precisely the characteristic Mott-insulator nature of these materials.
The success of pCCD can be attributed to its connection with seniority-zero methods, particularly perfect pairing and generalized valence bond approaches; \cite{cullen1996generalized,stein2014seniority,kutzelnigg2012separation,voorhis2000nonorthogonal,van2001connections,small2012fusion} which explains its ability to describe non-dynamic electronic correlation effects ~\cite{boguslawski2016analysis}.  However, to achieve quantitatively meaningful results, pCCD must be combined with an orbital optimization procedure. \cite{boguslawski2014efficient,henderson2014seniority}

It is known that the usual operators employed in quantum chemistry are real-valued in time-independent applications. Hence, the use of complex orbitals has been less explored in favor of real orbitals. Nevertheless, complex orbitals have attracted attention from the community due to the extra flexibility provided by the complex parameterization. \cite{fukutome1981unrestricted,scuseria:11jcp,jimenez-hoyos:12jcp,small2015restricted,jimenez2011generalized,song2024combiningcomplexconjugationtimereversal,stuber2003symmetry} Specifically, it has shown to be an efficient alternative to multiconfigurational methods to account for non-dynamic electronic correlation effects with single-determinant wave functions.\cite{jimenez2011generalized,song2024combiningcomplexconjugationtimereversal,stuber2003symmetry} However, complex-valued orbitals must be used carefully because they break symmetries among the spin-up ($\uparrow$) and spin-down ($\downarrow$) electrons trivially granted by real-valued orbitals. 

In the absence of spin-orbit coupling contributions to the Hamiltonian or external magnetic fields, the spin-with orbitals ($\chi_{p\uparrow}$ and $\chi_{p\downarrow}$) are typically constructed as the direct product between a spin function ($\uparrow$ or $\downarrow$) and a spatial function $\chi _p ({\br})=\sum _{\mu} c_{\mu p} \psi_\mu({\br})$ built as a linear combination of real atomic orbitals $\psi_\mu(\br)$, where the matrix $\bc$ gathers the molecular orbital coefficients. \rev{(Note that the convention chosen designates the natural orbitals by $\chi$ and the basis functions by $\psi$, whereas the opposite convention is often used in the quantum chemistry community.)} In general, real orbitals ($\bc \in \mathbb{R}$) are constructed to preserve fundamental symmetries such as spin symmetries ($\hS_z$ and $\hS^2$), complex conjugation ($\hcK$), and time-reversal symmetry ($\hTheta$). On the contrary, working with complex orbitals ($\bc \in \mathbb{C}$), one is forced to preserve either $\hS^2$ and $\hS_z$ (charge current wave in Fukutome's classification) or $\hTheta$ and $\hS_z$ (axial spin current wave in Fukutome's classification). \cite{fukutome1981unrestricted,stuber2003symmetry,jimenez2011generalized,small2015restricted} In the former case, the spatial part of the orbitals, $\chi_{p} (\br)$, for the spin-up and spin-down electrons is identical, which guarantees the correct value of $\expval*{\hS^2}$. In the latter case, the spin-up orbitals, $\chi_{p\uparrow} (\br)$, are related to the spin-down orbitals, $\chi_{p\downarrow} (\br)$, as
 \begin{equation}
 \begin{split}
 \hTheta\chi_{p\uparrow} (\br) & =
\hTheta\chi_{p} (\br) \mqty( 1 \\ 0 ) = 
 -\ii \;\sigma_y \; \hcK \chi_{p} (\br) \mqty( 1 \\ 0 ) 
 \\
   & = -\ii \mqty(
    0    & -\ii \\
     \ii   & 0
   ) \hcK\chi_{p} (\br) \mqty( 1 \\ 0 )  = \chi^* _{p}  (\br) \mqty( 0 \\ 1 )=\chi_{p\downarrow} (\br) ,
 \end{split}
 \label{eq:time_rev_def}
 \end{equation}
where $\ii=\sqrt{-1}$ and $\sigma_y$ is a Pauli matrix. This option preserves time-reversal symmetry, that is, $\hTheta\chi_{p\uparrow} (\br)  = \chi_{p\downarrow} (\br)$ and $\hTheta\chi_{p\downarrow} (\br) = - \chi_{p\uparrow} (\br)$, but not $\hS^2$. 

\rev{Before proceeding further, let us express the spin-with orbitals obtained by imposing time-reversal symmetry in matrix form as
\begin{equation}
 \begin{split}
 \mqty(
    \boldsymbol{\chi} &  {\bf 0} \\
    {\bf 0} & \boldsymbol{\chi}^*
   ) =
   \mqty(
    \bc    & {\bf 0} \\
    {\bf 0}   & \bc ^*
   ) \mqty(
    \boldsymbol{\psi} & {\bf 0}\\
    {\bf 0} & \boldsymbol{\psi}
   ).
 \end{split}
 \label{eq:c_spinors}
 \end{equation}
As previously mentioned, this convention also preserves $\hS_z$ (i.e., it fixes the number of spin-up and spin-down electrons).
However, the axis chosen for the quantization is irrelevant for Hamiltonians that do not account for spin-orbit coupling effects or external magnetic fields. Therefore, rather than working in the eigenbasis of $\hS_z$ we may equally well choose to work in the eigenbasis of $\hS_y$ whose eigenstates ($\chi^{y} _{p\uparrow}$ and $\chi^{y} _{p\downarrow}$) are given as a linear combination of the spin-up and spin-down eigenbasis of $\hS_z$, i.e.,
\begin{align}
    \chi^{y} _{p\uparrow}(\br) & = \frac{\chi_{p\uparrow} ({\br})+\ii \chi_{p\downarrow} ({\br})}{\sqrt{2}},
    &
    \chi^{y} _{p\downarrow}(\br) & = \frac{\ii \chi_{p\uparrow} ({\br})+ \chi_{p\downarrow} ({\br}) }{\sqrt{2}}.
\end{align}
If we also rotate the basis functions to the eigenbasis of $\hS_y$ 
\begin{align}
    \psi^{y} _{p\uparrow}(\br) & = \frac{\psi_{p\uparrow} ({\br})+\ii \psi_{p\downarrow} ({\br})}{\sqrt{2}},
    &
    \psi^{y} _{p\downarrow}(\br) & = \frac{\ii \psi_{p\uparrow} ({\br})+ \psi_{p\downarrow} ({\br}) }{\sqrt{2}}
\end{align}
we obtain: 
\begin{equation}
 \begin{split}
   \boldsymbol{\chi}^y =
 \frac{1}{\sqrt{2}}
 \mqty(
    {\bf 1}    & \ii{\bf 1} \\
    \ii{\bf 1}   & {\bf 1})
 \mqty(
      \boldsymbol{\chi} &  {\bf 0} \\
    {\bf 0} & \boldsymbol{\chi}^*
   ) =
    \frac{1}{2}
     \mqty(
    {\bf 1}    & \ii{\bf 1} \\
    \ii{\bf 1}   & {\bf 1})
   \mqty(
    \bc    & {\bf 0} \\
    {\bf 0}   & \bc ^*
   )
    \mqty(
    {\bf 1}    & -\ii{\bf 1} \\
    -\ii{\bf 1}   & {\bf 1})
   \mqty(
    \boldsymbol{\psi}^y & {\bf 0}\\
    {\bf 0} & \boldsymbol{\psi}^y
   )\\
   = \frac{1}{2} \mqty(
    \bc + \bc^*    & -\ii\bc + \ii\bc^* \\
    \ii\bc - \ii\bc^*   &\bc + \bc^* 
   )\mqty(
    \boldsymbol{\psi}^y & {\bf 0}\\
    {\bf 0} & \boldsymbol{\psi}^y
   ) =\mqty(
    \Re(\bc)    & \Im (\bc) \\
    -\Im (\bc)   & \Re(\bc) 
   )\mqty(
    \boldsymbol{\psi}^y & {\bf 0}\\
    {\bf 0} & \boldsymbol{\psi}^y
   ).
 \end{split}
 \label{eq:c_z_to_y}
 \end{equation}
This choice of eigenbasis therefore leads to real coefficients at the expense of introducing 2-component spinors. When applied to the HF approximation, the latter corresponds to the real-paired generalized HF method, which is defined in this work as a particular case of the paired-generalized HF method~\cite{stuber2003symmetry} (see Table 1 in Ref.~\onlinecite{jimenez2011generalized}). In paired-generalized HF, the coefficients of the 2-component spinors are complex, and only time-reversal symmetry is preserved. However, as shown above, by additionally imposing real algebra for the 2-component spinors, $\hS _y$ is also preserved. In summary, working with the real-paired generalized HF method is equivalent to working with complex orbitals. In both cases time-reversal symmetry is preserved, but the former preserves $\hS _y$ while the latter preserves $\hS _z$  (and is named paired UHF in Stuber--Paldus designation \cite{stuber2003symmetry}). 
}

In this work, we have preferred to work with complex orbitals (instead of 2-component real spinors) and preserve $\hS_z$. Also, we have enforced the spin-up and spin-down orbitals to be related by complex conjugation of the spatial part to preserve time-reversal symmetry. Thus, deviations from the physical $\expval*{\hS^2}$ value (i.e., spin contamination) might occur. This selection is motivated by three reasons: (i) time-reversal symmetry is typically imposed in codes that can deal with complex orbitals/one-body states for extended systems (e.g. \texttt{ABINIT} \cite{gonze2020abinit} or \texttt{QUANTUM ESPRESSO}, \cite{giannozzi2009quantum,giannozzi2017advanced} see also Appendix B in Ref.~\onlinecite{denawi2023g}), (ii) the simplification in the RDMFT and pCCD equations facilitates their extension to complex-valued orbitals, and (iii) it corresponds to the correct non-relativistic limit when time-reversal symmetry is employed to build the 4-component spinors that are employed in the solution of the Dirac-Coulomb/Coulomb-Gaunt Hamiltonians. \cite{dyall2007introduction} 

To gain further insights into this limit, let us first mention that the relativistic 4-component spinors are complex-valued and can be chosen to preserve time-reversal symmetry in the absence of external magnetic fields (i.e., forming Kramers' pairs \cite{kramers1930theorie,aucar1995operator,buvcinsky2015spin,buvcinsky2016additivity,komorovsky2016new}). They are usually expanded in two distinct basis sets, one for the upper (large) components and one for the lower (small) components of the Dirac wave function. For better comparison with non-relativistic basis set expansions, it is therefore convenient to perform an exact transformation to the 2-spinor (X2C) form. \cite{jensen:rehe2005, kutzelnigg:jcp2005, iliavs2007infinite} The expansion coefficients for these X2C-spinors are complex and the preservation of time-reversal symmetry is easily visible when they are written in matrix form as 
\begin{equation}
\begin{split}
\mqty(
   \bc _{L\uparrow}    & \bc _{L\downarrow} \\
   -\bc _{L\downarrow}^*   & \bc _{L\uparrow} ^* 
  ). 
\end{split}
\end{equation}
This corresponds to the torsional spin current wave (TSCW) in Fukutome's classification. \cite{fukutome1981unrestricted,jimenez2011generalized} In the X2C form, it is possible to use the Dirac identity to remove spin-orbit coupling terms from the Hamiltonian. This simplifies the matrix representation of the Hamiltonian, making it real-symmetric, and thus, one typically proceeds using real coefficients, as seen in Eq.~\ref{eq:c_z_to_y}, \rev{i.e.
 \begin{equation}
 \begin{split}
\mqty(
   \bc _{L\uparrow}    & \bc _{L\downarrow} \\
   -\bc _{L\downarrow}^*   & \bc _{L\uparrow} ^* 
  ) \rightarrow  \mqty(
    \Re(\bc)    & \Im (\bc) \\
    -\Im (\bc)   & \Re(\bc) 
   ). 
 \end{split}
 \label{eq:c_real_2c_relation}
 \end{equation}
As previously mentioned, working with real 2-spinors that preserve time-reversal symmetry is equivalent to using complex spatial orbitals [Eqs.~\eqref{eq:c_spinors} and \eqref{eq:c_z_to_y}] that preserve this symmetry, where only the spin quantization axis is changed (i.e.~they correspond to either eigenfunctions of $\hS _y$ or $\hS _z$, respectively). However, employing complex spatial orbitals (or real 2-spinors) involves a reduction of the variational freedom with respect to complex 2-spinors. Despite this, complex spatial orbitals still retain more flexibility than real spatial orbitals.} Hence, real spatial orbitals are only obtained during orbital optimization if the additional freedom provided by the complex coefficients does not lead to a lower energy solution according to the variational principle (see Sec.~\ref{sec:beh2_n2}).
 
The analysis of the consequences of using complex algebra and imposing time-reversal symmetry on the actual universal RDMFT functional (and approximate functionals built from a multi-configurational self-consistent field expansion) can be found in a previous work~\cite{liebert2023refining}. In contrast, here we examine the role of complex orbitals in extending the applicability of the most recently developed RDMFT functionals and pCCD to systems described by complex single-particle states. In Sec.~\ref{sec:theo}, we briefly introduce RDMFT (Sec.~\ref{sec:rdmft}), pCCD (Sec.~\ref{sec:pCCD}), and the orbital optimization procedure proposed by Ugalde and Piris (Sec.~\ref{sec:orb_opt}), \cite{piris2009iterative} which is currently applied in most RDMFT calculations. Then, in Sec.~\ref{sec:theo_res}, we discuss the effect of time-reversal symmetry on the RDMFT and pCCD energy expressions and its impact on the orbital optimization procedure. Finally, in Sec.~\ref{sec:prac_res}, we present some numerical examples that illustrate the consequences of using complex-valued orbital coefficients and time-reversal symmetry in practical calculations for spin-compensated systems.  Our conclusions are drawn in Sec.~\ref{sec:ccl}. Unless otherwise stated, atomic units are used throughout.

\section{Theoretical background}
\label{sec:theo}

\subsection{Reduced Density Matrix Functional Theory}
\label{sec:rdmft}

In 1975, Gilbert \cite{gilbert:75prb} proposed an extension of the Hohenberg-Kohn theorem \cite{hohenberg:64pr} for non-local external potentials, which introduces the energy as a functional of the first-order reduced density matrix (1RDM) $\gamma$. It generalizes the functional based on the electronic density $\rho$ that is employed in density-functional theory.\cite{burke:07book} A compact representation of $\gamma$ is obtained by expressing it in the natural orbital basis
\begin{equation}
    \gamma(\br,\br')=\sum _p \sum _{\sigma=\uparrow,\downarrow} n_{p\sigma} \chi_{p\sigma}^{*} (\br) \chi_{p\sigma} (\br'),
\end{equation} 
where $n_{p\sigma} \in \qty[0,1]$ is the occupation number associated with the natural spin-orbital $\chi_{p\sigma}$. For $\br=\br'$, it reduces to the electron density, that is, $\rho(\br)=\gamma(\br,\br)$. In practical realizations of RDMFT, the matrix elements of the second-order reduced density matrix (2RDM) ${}^2D^{\sigma \sigma'}_{pq,rs}=\mel{ \Psi }{ \cre{c}{p\sigma} \cre{c}{q\sigma'} \anh{c}{s\sigma'} \anh{c}{r\sigma} }{ \Psi } $ (with $\sigma' = \ \uparrow$ or $\downarrow$) are expressed as functions of the occupation numbers and the natural orbitals are employed to compute the two-electron repulsion integrals. Here, $\cre{c}{p\sigma}$ ($\anh{c}{p\sigma}$) is the usual creation (annihilation) operator and $\Psi$ is the exact $N$-electron wave function. In the most basic RDMFT approximations,\cite{rodriguez:17pccp_2,schilling2019diverging} the 2RDM elements of the opposite-spin ($\sigma \neq \sigma'$) and same-spin ($\sigma = \sigma'$) blocks read
\begin{subequations}
\begin{align}
{}^2 D^{\sigma \sigma'}_{pq,rs} & = \frac{n_{p\sigma} n_{q\sigma'}}{2} \delta_{pr}\delta_{qs},
 \label{eq:jk2rdm_ssp}
 \\
{}^2 D^{\sigma \sigma} _{pq,rs} & = \frac{n_{p\sigma} n_{q\sigma}}{2} \delta_{pr}\delta_{qs}-\frac{ f(n_{p\sigma},n_{q\sigma })}{2}\delta_{ps}\delta_{qr},
 \label{eq:jk2rdm_ss}
\end{align}
\end{subequations}
with $f(n_{p\sigma},n_{q\sigma})$ being a simple function of the occupations numbers. For example, $f(n_{p\sigma},n_{q\sigma})=\sqrt{n_{p\sigma}n_{q\sigma}}$ in the M\"uller functional~\cite{muller:84pl,buijse:02mp,buijse:91thesis} and $f(n_{p\sigma},n_{q\sigma})=(n_{p\sigma}n_{q\sigma})^\alpha$ with $\alpha \in \mathbb{R}^+$ in the power functional~\cite{sharma:08pow,cioslowski2000:power}. Note that the elements ${}^2D^{\sigma \sigma'}_{pq,pq} = n_{p\sigma} n_{q\sigma}/2$ correspond to Hartree contributions while the terms $ {}^2D^{\sigma \sigma}_{pq,qp} = -f(n_{p\sigma},n_{q\sigma})/2$ are modified exchange contributions that account for electronic correlation effects. Hence, these approximations are $JK$-only functionals \cite{rodriguez:17pccp_2} because only Hartree ($J$) and exchange ($K$) integrals are required in the evaluation of the electronic energy. However, more advanced RDMFT functionals based on the reconstruction of the second-order cumulant matrix, such as PNOF$i$~\cite{piris:13ijqc,piris:13jcp1,piris:17sub} ($i=5$, $6$, and $7$) and GNOF, \cite{piris:21gnof} include the additional $L$ integrals~\cite{piris:99jmc}, defined below. Thus, they are usually referred to as $JKL$-only approximations. For spin-compensated systems ($n_p=n_{p\uparrow}=n_{p\downarrow}$), the electronic energy functional of PNOF$i$ and GNOF takes the following form
\begin{equation}
E^{\text{RDMFT}}=2\sum _p n_p h_{pp} + \sum _{pq} \qty( {}^2 D_{pq,pq} J_{pq} + {}^2 D_{pq,qp} K_{pq}+ {}^2 D_{pp,qq} L_{pq} ), 
 \label{eq:E_jkl}
\end{equation}
where the one-electron integrals are
\begin{equation}
    h_{pq} = \int \dd\br \chi^*_p (\br) \hh \chi_q (\br),
\end{equation}
with $ \hh = -\nabla^2_{\br}/2 + v_\text{ext}(\br)$ being the (one-electron) core Hamiltonian and $v_\text{ext}(\br)$ is the external potential. The various types of two-electron integrals
\begin{align}
    J_{pq} &= \braket{pq}{pq}, &
    K_{pq} &= \braket{pq}{qp}, &
    L_{pq} &= \braket{pp}{qq},
\end{align}
are expressed in terms of the spatial part of the natural orbital basis,
\begin{equation}
    \braket{pq}{rs} = \iint \dd\br\dd\br' \frac{\chi_p ^* (\br)\chi_q ^* (\br')\chi_r (\br)\chi_s  (\br') }{\abs{\br-\br'}}.
\end{equation}
Note that, here, we have employed the spin-summed 2RDM elements, i.e., ${}^2 D_{pq,rs} = \sum _{\sigma \sigma'} {}^2D^{\sigma \sigma'}_{pq,rs}$. It is worth mentioning that L\"owdin normalization is used throughout this work, that is, $\Tr\qty[{}^2{\boldsymbol{D}} ]=N(N-1)/2$ with $N$ being the number of electrons. 

The energy contribution involving $L$ integrals arises from the interaction of opposite-spin electrons, i.e.
\begin{equation}
    \sum _{pq} {}^2 D_{pp,qq} L_{pq} = \sum_{pq} \sum _{\sigma \neq \sigma'} {}^2 D^{\sigma \sigma'}_{pp,qq} L_{pq}, 
\end{equation}
because the same-spin contributions cancel due to the Pauli exclusion principle (i.e., ${}^2 D^{\sigma \sigma}_{pp,qq} =0$). For real orbitals, it is easy to verify that $L_{pq}=K_{pq}$. Thus, the last term in Eq.~\ref{eq:E_jkl} is usually combined with the second term and the electronic energy is written using only $J$ and $K$ integrals. However, the $K$ and $L$ integrals differ for complex orbitals unless one imposes time-reversal symmetry (see below). It is easy to show that $J$ and $K$ integrals are real even for complex orbitals. On the contrary, $L$ integrals are complex-valued in general. 

\subsection{Coupled Cluster With Paired Doubles}
\label{sec:pCCD}

The pCCD approximation belongs to the coupled-cluster (CC) family of methods, which aims to go beyond the single-determinant wave function $\ket{0}$ to describe the many-body state using the wave operator $e^{\hT}$ to produce excited determinants from the reference determinant. In pCCD, for spin-compensated systems, the many-body wave function is expressed as $\ket{\Psi} = e^{\hT} \ket{0}$, where the excitation operator is restricted to paired double excitations \cite{limacher2013new} 
\begin{equation}
\hT = \sum _i ^{N/2} \sum _{a=N/2+1} ^{M} t_i ^a \cre{c}{a\uparrow}\cre{c}{a\downarrow} \anh{c}{i\downarrow}\anh{c}{i\uparrow},
\label{eq:T_amp}
\end{equation}
 where $M$ is the total number of spatial orbitals and the $t_{i}^a$'s are the so-called amplitudes that are optimized by solving the so-called amplitude (or residual) equations
 \begin{equation}
     \label{eq:t_amp}
 \begin{split}
     0&=\mel{0}{ \cre{c}{i\uparrow}\cre{c}{i\downarrow} \anh{c}{a\downarrow}\anh{c}{a\uparrow} \stH }{0} 
     \\
     &= L_{ai}+2\qty(f_{a} ^a - f_i ^i - \sum_j L_{ja} t_{j} ^a - \sum_b L_{ib} t_{i} ^b ) t_{i} ^a 
     \\
     &-2 \qty( 2 J_{ia} -K_{ia}-L_{ia} t_{i} ^a )  t_{i} ^a 
     +\sum _b L_{ba}  t_{i} ^b + \sum_j L_{ji} t_{j} ^a + \sum_{jb} L_{jb} t_{j} ^a  t_{i} ^b,
 \end{split}
 \end{equation}
 where $\stH = e^{-\hT} \hH e^{\hT}$ is the similarity-transformed Hamiltonian, $f_p ^q$ are the Fock matrix elements evaluated with $\ket{0}$, and $i$ and $j$ ($a$ and $b$) refer to occupied (virtual) orbitals with respect to $\ket{0}$. The amplitude equations, which are quadratic in $t$, can be solved in $M^3$ computational cost by building the intermediate $y_{i} ^j = \sum_b L_{jb} t_{i} ^b $. 

To perform orbital optimizations~\cite{boguslawski2014efficient,boguslawski2014nonvariational,boguslawski2014projected,henderson2014seniority}, let us introduce the pCCD energy functional as
 \begin{equation}
  E^{\text{pCCD}}=\mel{ \mathcal{L} }{ \stH }{0},  
  \label{eq:E_pccd1}
 \end{equation}
where $\bra{\mathcal{L}}= \bra{0}  (1+ \hZ)$ is the left eigenvector of $\stH$ and $\hZ = \sum_{ia} z_{i}^a \cre{c}{i\uparrow}\cre{c}{i\downarrow} \anh{c}{a\downarrow}\anh{c}{a\uparrow}$ is a deexcitation operator. \rev{(Note that this energy functional corresponds to the one used in the Lagrangian formulation that leads to the well-known $\Lambda$-equations. \cite{helgaker:00book} However, the notations chosen in this work coincide with the ones used in Ref.~\onlinecite{henderson2014seniority}.)}
The stationary conditions $\pdv*{E^\text{pCCD}}{z_i^a} =0$ yield the $t$-amplitude equations [see Eq.~\ref{eq:t_amp}] while the additional conditions $\pdv*{E^\text{pCCD}}{t_i^a} =0$ allows us to write the (linear) residual equations for the left amplitudes $\{ z_{i}^a \}$ as
 \begin{equation}
 \label{eq:z_amp}
 \begin{split}
     0&= L_{ia}+2\qty(f_{a} ^a - f_i ^i - \sum_j L_{ja} t_{j} ^a - \sum_b L_{ib} t_{i} ^b ) z_{i} ^a 
     -2\qty( 2 J_{ia} -K_{ia}- 2 L_{ia} t_{i} ^a )  z_{i} ^a 
     \\
     &+\sum _b L_{ab}  z_{i} ^b + \sum_j L_{ij} z_{j} ^a + \sum_{jb} t_{j} ^b \left(L_{ib}z_{j} ^a + L_{ja} z_{i} ^b \right)
     -2 L_{ia}  \qty( \sum _j z_j ^a t_j ^a + \sum _b z_i ^b t_i ^b ). 
 \end{split}
 \end{equation}
 Next, with the aid of the $t$- and $z$-amplitude equations, one can easily compute the 1RDM, which is diagonal within the pCCD approximation \cite{henderson2014seniority}   
 \begin{equation}
  {}^1D^{\sigma} _{p,q}  =\mel{0}{(1+ \hZ ) e^{-\hT}\cre{c}{p\sigma}\anh{c}{q\sigma}e^{\hT}}{0} \delta_{pq} = n_{p\sigma} \delta_{pq},
  \label{eq:pccd_1rdm}
 \end{equation}
 and directly linked to the occupation numbers that can be written as $n_{i\sigma} = (1-x_i^i)$ and $n_{a\sigma} = x_a^a$ with $x_i ^j = \sum _a t_i ^a z_j ^a$ and $x_a ^b = \sum _i t_i ^b z_i ^a$. Similarly, we may write the matrix elements of the spin-summed 2RDM 
\begin{equation}
^2D_{pq,rs}=\sum_{\sigma\sigma'} \mel{0}{(1+ \hZ) e^{-\hT}\cre{c}{p\sigma}\cre{c}{q\sigma'} \anh{c}{s\sigma'}\anh{c}{r\sigma} e^{\hT}}{0},
\label{eq:pccd_2rdm}
\end{equation}
as
\begin{subequations}
\begin{align}
    ^2D_{ii,jj} &= x_i ^j + \delta_{ij} (1-2 x_i ^i), \\
    ^2D_{ii,aa} &= t_i ^a + x_i ^a -2 t_i ^a(x_a ^a + x_i ^i - t_i ^a z_i ^a), \\
    ^2D_{aa,ii} &= z_i ^a,  \\
    ^2D_{aa,bb} &= x_a ^b,  \\
    ^2D_{ij,ij} &= 2(1-x_i^i x_j^j) + \delta _{ij} 3 (x_i ^i -1), \\
    ^2D_{ia,ia} &={}^2D_{ai,ai}=2(x_a ^a -t_i ^a z_i^a),\\
    ^2D_{ab,ab} &=\delta_{ab} x_a ^a, \\
    ^2D_{pq,qp} &= \frac{{}^2D_{pq,pq}}{2} \qfor p \neq q,
\end{align}
\label{eq:pccd_2rdm_me}
\end{subequations}
where we have employed the additional intermediate $x_i ^a = \sum_{jb} t_i ^b t_j ^a z_j^b$. 

Noticing that the non-zero spin-summed 2RDM elements are the same as in PNOF$i$/GNOF and that the 1RDM is expressed in its diagonal representation, we recognize that the pCCD energy can also be written as
\begin{equation}
E^{\text{pCCD}}=2\sum _p n_p h_{pp} + \sum _{pq} \left( {}^2 D_{pq,pq} J_{pq} + {}^2 D_{pq,qp} K_{pq}+ {}^2 D_{pp,qq} L_{pq} \right),
 \label{eq:E_jkl_pccd}
\end{equation}
which exactly matches Eq.~\eqref{eq:E_jkl}. Once more, the contributions involving the time-reversal integrals $L$ arise from the interaction of opposite-spin electrons as in RDMFT approximations. \rev{It is worth mentioning that the pCCD energy for a given (fixed) set of molecular orbitals can be written as 
\begin{equation}
  E^{\text{pCCD}}=\mel{0}{\hH}{0} + \sum ^{N/2} _{i} \sum ^M _{a=N/2+1} t_i ^a  L_ {ia},
\end{equation}
where only the reference wavefunction $\ket{0}$, the $t_i^a$ amplitudes, and the $L_{ia}$ integrals are needed. However, to perform orbital optimizations within coupled-cluster methods, it is necessary to define a variational functional \cite{helgaker:00book} (such as the one given in Eq.~\eqref{eq:E_pccd1}). Then, all the possible 1RDM and 2RDM elements have to be built because they enter into the definition of the gradient (and the Hessian) of the energy functional with respect to orbital rotations (see below). Interestingly, the 1RDM and 2RDM elements for the pCCD method (given by Eqs.~\eqref{eq:pccd_1rdm}-\eqref{eq:pccd_2rdm_me}) reproduce the structure of seniority--zero wavefunctions~\cite{richardson:95jacs} due to the type of excitations involved by this method on the reference wavefunction $\ket{0}$.} Finally, let us remark that the $L$ integrals are also present in the definition of the $t$ and $z$ amplitudes [see Eqs.~\eqref{eq:t_amp} and \eqref{eq:z_amp}]. Therefore, if one relies on complex-valued orbitals, the resulting amplitudes are also complex in general.    

\subsection{Orbital Optimization}
\label{sec:orb_opt}

It is well-documented~\cite{lathiotakis2010size,boguslawski2014efficient,boguslawski2014nonvariational,boguslawski2014projected,henderson2014seniority,cioslowski:15jcp} that the $JKL$-only RDMFT approximations and pCCD are not invariant with respect to orbital rotations (even for the occupied-occupied and virtual-virtual blocks). Thus, orbital optimization is required to correctly describe the electronic structure, especially in spatial regions where non-dynamic electronic correlation effects are dominant. Since the electronic energies of $JKL$-only RDMFT approximations and pCCD have the same form, as readily seen in Eqs.~\eqref{eq:E_jkl} and \eqref{eq:E_jkl_pccd}, the same orbital optimization machinery can be employed. Here, we consider first the algorithm proposed by Piris and Ugalde \cite{piris2009iterative}, which optimizes the occupation numbers and the orbitals in a two-step iterative process (i.e., by neglecting the coupling between occupations and orbitals). 

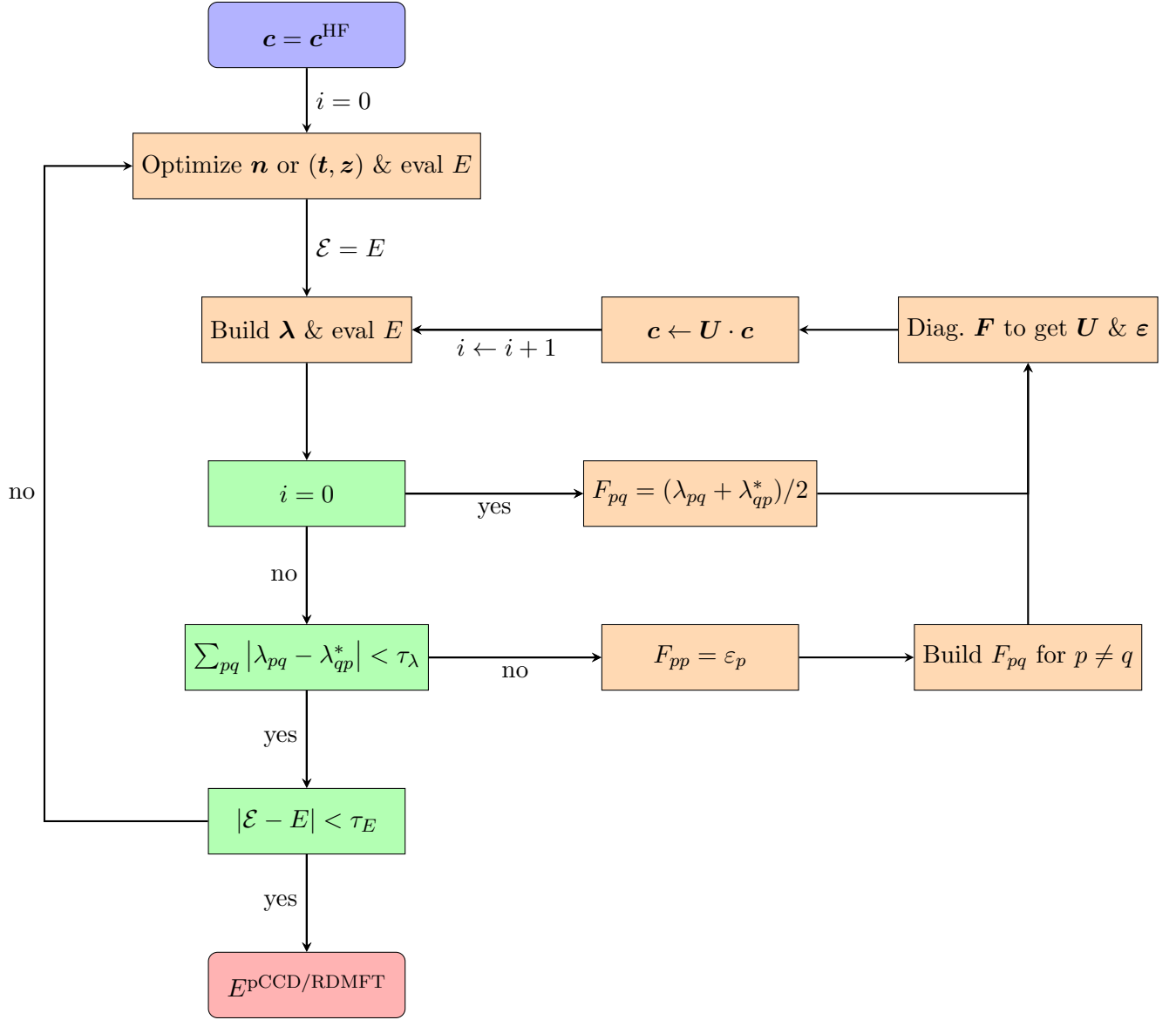
\begin{figure}[H]
%\hspace{-1cm}
\begin{tikzpicture}[node distance=2.0cm]
%\node (start) [start] {pCCD or RMDFT};
\node (in1) [io] {$\bc = \bc^{\textrm{HF}}$};
\node (pro1) [process, below of=in1] {Optimize $\boldsymbol{n}$ or ($\boldsymbol{t},\boldsymbol{z}$) \& eval $E$};
\node (pro2) [process, below of=pro1, yshift=-0.5cm] {Build $\boldsymbol{\lambda}$ \& eval $E$};
\node (dec1) [decision, below of=pro2, yshift=-0.5cm] {$i=0$};
\node (pro3b) [process, right of=pro2, xshift=4cm] {${\bc} \leftarrow \boldsymbol{U} \cdot {\bc} $};
\node (pro3) [process, right of=dec1, xshift=4cm] {$F_{pq} =(\lambda_{pq}+\lambda^*_{qp})/2$};
\node (dec2) [decision, below of=dec1, yshift=-0.5cm] {$\sum_{pq}\abs{\lambda_{pq}-\lambda^*_{qp}}<\tau_\lambda$};
\node (pro4) [process, right of=dec2, xshift=4cm] {$F_{pp}=\varepsilon_p$};
\node (pro3c) [process, right of=pro3b, xshift=3cm] {Diag.~$\boldsymbol{F}$ to get $\boldsymbol{U}$ \& $\boldsymbol{\varepsilon}$};
\node (pro5) [process, right of=pro4, xshift=3cm] {Build $F_{pq}$ for $p\neq q$};
\node (dec3) [decision, below of=dec2, yshift=-0.5cm] {$\abs{\mathcal{E}-E}<\tau_E$};
\node (stop) [stop, below of=dec3, yshift=-0.5cm] {$E^{\textrm{pCCD/RDMFT}}$};

%\foreach \Point in {(-4,-4)}{
%    \node(point) at \Point {\tiny{\textbullet}};
%}

%\draw [arrow] (start) -- (in1);
\draw [arrow] (in1) -- node[anchor=west] {$i = 0$} (pro1);
\draw [arrow] (pro1) --  node[anchor=west] {$\mathcal{E} = E$} (pro2);
\draw [arrow] (pro2) -- (dec1);
\draw [arrow] (pro3b) -- (pro2);
\draw [arrow] (dec1) -- (pro3);
\draw [arrow] (pro3) -| (pro3c);
\draw [arrow] (dec1) -- (dec2);
\draw [arrow] (dec2) -- (dec3);
\draw [arrow] (dec3) -- (stop);
\draw [arrow] (dec2) -- (pro4);
\draw [arrow] (pro4) -- (pro5);
\draw [arrow] (pro3c) -- (pro3b);
\draw [arrow] (pro5) -- (pro3c);
%\draw [arrow] (point) -- (pro1);
\draw [arrow] (dec1) -- node[anchor=north] {yes} (pro3);
\draw [arrow] (dec2) -- node[anchor=north] {no} (pro4);
\draw [arrow] (dec1) -- node[anchor=east] {no} (dec2);
\draw [arrow] (dec2) -- node[anchor=east] {yes} (dec3);
%\draw [arrow] (dec3) -| node[anchor=east] {no} (point);
\draw [arrow] (dec3) -- node[anchor=east] {yes} (stop);
\draw [arrow] (pro3b) -- node[anchor=north] {$i \leftarrow i+1$} (pro2);
\draw[arrow] (dec3.west) -- (-4,-12) -- node[anchor=east] {no} (-4,-2) -- (pro1.west);
\end{tikzpicture}
    \caption{Orbital optimization procedure based on the Piris-Ugalde algorithm \cite{piris2009iterative} employed in pCCD and RDMFT. 
    The matrix $\bc$ gathers the spin-up natural orbital coefficients, $\bn$ is the set of occupations in RDMFT, $\bt$ and $\bz$ contain the right and left amplitudes in pCCD, respectively, while the matrices $\boldsymbol{U}$ and $\boldsymbol{\varepsilon}$ gathers the eigenvectors and eigenvalues of $\bF$, respectively. Two thresholds are introduced. One of them, $\tau_\lambda$, controls the deviation from Hermiticity of $\boldsymbol{\lambda}$ while the other, $\tau_E$, monitors the energy convergence.}
    \label{fig:scf_algo}
\end{figure}

The central quantity of the Piris-Ugalde constrained optimization procedure is the following Lagrangian which reads, for fixed occupation numbers and spin-summed 2RDM elements,
\begin{equation}
    \Omega = E^{\text{pCCD}/\text{RDMFT}} -\sum_{pq} \lambda_{pq} \qty(\braket{ \chi _p}{\chi_q} - \delta_{pq}), 
\end{equation}
where $\braket{ \chi_p}{\chi_q}$ is the overlap of the spatial part of the natural orbitals and the $\lambda_{pq}$'s are Lagrange multipliers which enforce the orthogonality of the natural orbitals during the optimization process. The Lagrangian $\Omega$ must be stationary with respect to the orbital variations, which is enforced by the following condition:
\begin{equation}
\begin{split}
    \pdv{E^{\text{pCCD}/\text{RDMFT}}}{\chi ^* _p (\br)} &= 2 n_p \hat{h} \chi_p (\br)+ 2 \sum_{r} {}^2D_{rr,rr} \pdv{J_{rr}}{\chi ^* _p (\br)}   
    \\
    & + 2\sum_{r \neq s} \qty[  {}^2D_{sr,sr} \pdv{J_{sr}}{\chi ^* _p (\br)} + {}^2D_{sr,rs} \pdv{K_{sr}}{\chi ^* _p (\br)} + {}^2D_{ss,rr} \pdv{L_{sr}}{\chi ^* _p (\br)} ]
    \\
    &= \sum_s \lambda_{ps} \chi_s (\br).
\end{split}
\end{equation}
Multiplying from the left by $\chi_q ^* (\br)$ and integrating over the spatial coordinates leads to
\begin{equation}
\begin{split}
    \lambda_{pq} 
    & = 2 \qty( n_p h_{qp} +{}^2D_{pp,pp} \braket{qp}{pp} ) 
    \\
    & + 2\sum_{r \neq p} \qty(  {}^2D_{pr,pr} \braket{qr}{pr} + {}^2D_{pr,rp} \braket{qr}{rp} + {}^2D_{pp,rr} \braket{qp}{rr} ). 
\end{split}
\end{equation}
Then, imposing the Hermiticity of the matrix $\boldsymbol{\lambda}$ at the stationary solution (i.e., $\lambda_{pq} = \lambda_{qp}^*$), the auxiliary Hermitian matrix $\boldsymbol{F}$, with elements 
\begin{equation}
F_{pq}=\begin{cases}
\lambda_{qp}-\lambda_{pq}^*  \qfor & p>q, \\
\lambda_{pq}^* -\lambda_{qp} \qfor & p<q, \\
\end{cases}
\end{equation} 
is built to perform orbital rotations (see Fig.~\ref{fig:scf_algo} for more details). 
The diagonal elements of $\boldsymbol{\lambda}$ read
\begin{equation}
    \lambda_{pp} = 2 \qty( n_p h_{pp} +{}^2D_{pp,pp} J_{pp} ) 
    +2\sum_{r \neq p} \qty(  {}^2D_{pr,pr} J_{pr}+ {}^2D_{pr,rp} K_{pr} + {}^2D_{pp,rr} L_{pr} ).  
\end{equation}
Therefore, for real elements ${}^2D_{pp,qq}$ that satisfy ${}^2D_{pp,qq}={}^2D_{qq,pp}$ (as it happens in RDMFT approximations), the diagonal elements of $\boldsymbol{F}$ are zero for real orbitals, i.e.,
\begin{equation}
    \lambda_{pp}-\lambda_{pp} ^* = 2\sum_{r \neq p} {}^2D_{pp,rr} \qty( L_{pr}- L_{rp} ) = 4\sum_{r \neq p} {}^2D_{pp,rr} \Im L_{pr} ,  
\label{eq:lpp_lpp}
\end{equation}
where $ \Im L_{pr} $ is the imaginary part of the matrix element $L_{pr}$ (which is zero for real orbitals). 
Hence, $\lambda_{pp}-\lambda_{pp}^*=0$. Consequently, it has been proposed to define the initial elements of the Fock matrix as $F_{pq} = (\lambda_{pq}+\lambda_{qp}^*)/2$. 
Then, the iterative construction and diagonalization of $\bF$ for fixed occupation numbers and 2RDM elements produce a set of optimal orbitals. 
Let us mention that, at a given iteration, the eigenvalues $\boldsymbol{\varepsilon}$ obtained from the diagonalization of $\bF$   are used as its diagonal elements for the next iteration (see Fig.~\ref{fig:scf_algo}).

The orbital optimization algorithm is preceded by the optimization of the occupation numbers, $\bn$, in RDMFT approximations or of the sets of amplitudes, $\bt$ and $\bz$, in pCCD. Therefore, the optimization procedure consists of an algorithm composed of two uncoupled steps that are controlled by two thresholds, $\tau_\lambda$ and $\tau_E$, that monitor the deviation from Hermiticity of $\boldsymbol{\lambda}$ and the energy convergence, respectively (see Fig.~\ref{fig:scf_algo}).

Other algorithms for the optimization of the orbitals employ the unitary matrix $e^{\bkappa}$ to perform orbital rotations, which is built as the exponential of an anti-Hermitian matrix $\bkappa$ with elements $\kappa_{pq} \in \mathbb{C}$ and $\kappa_{pq}=-\kappa^*_{qp}$~\rev{(see, for example, Refs.~\onlinecite{boguslawski2014efficient,boguslawski2014nonvariational,boguslawski2014projected,henderson2014seniority} in the case of pCCD)}. Additionally, one may introduce the corresponding rotation operator $e^{\hkappa}$ that is applied to the wave function to obtain the transformed wave function $\ket*{\tilde{\Psi}} =e^{\hat{\kappa}}\ket{\Psi}$ built with from these rotated orbitals, where \cite{helgaker:00book}
\begin{equation}
\begin{split}
\hkappa &= \sum_{pq} \sum_{\sigma} \kappa_{pq} \cre{c}{p\sigma}\anh{c}{q\sigma} \\
&=\sum_{p} \sum_{\sigma} \ii\Im \kappa_{pp}  \cre{c}{p\sigma}\anh{c}{p\sigma} + \sum_{p>q} \sum_{\sigma} \qty(\Re \kappa_{pq} + \ii\Im \kappa_{pq}) \qty(\cre{c}{p\sigma}\anh{c}{q\sigma}-\cre{c}{q\sigma}\anh{c}{p\sigma}),
\end{split} 
\end{equation} 
with $\Re \kappa_{pp}=0$.
Assuming that $\braket{\Psi} = 1$, the energy of  $\ket*{\tilde{\Psi}}$ can be written as 
\begin{equation}
    E(\bkappa) 
    = \frac{\mel*{\tilde{\Psi}}{\hH}{\tilde{\Psi}}}{\braket*{\tilde{\Psi}}{\tilde{\Psi}}} 
    = \mel{\Psi}{e^{-\hkappa} \hH  e^{\hkappa}}{\Psi},
\end{equation}
that has to be made stationary with respect to the orbital rotation parameters $\kappa_{pq}$, i.e., $\eval{\pdv*{E(\bkappa)}{\kappa_{pq}}}_{\bkappa=\bO} = 0$, for each orbital pair.

Employing the Baker–Campbell–Hausdorff formula \cite{helgaker:00book}
\begin{equation}
     E(\bkappa)
     = \mel{\Psi}{\hH}{\Psi} 
     + \mel{\Psi}{\comm*{\hH}{\hkappa}}{\Psi}
     + \frac{1}{2}\mel{\Psi}{ \comm*{\comm*{\hH}{\hkappa}}{\hkappa}}{\Psi} 
     + \cdots,
\end{equation}
and introducing the elements of the gradient 
\begin{equation} \label{eq:gradient_elements}
 g_{pq}
 = {\eval{\pdv{E(\bkappa)}{\kappa_{pq}}}_{\bkappa=\bO}}
 = \frac{1}{2}\qty(\pdv{}{\Re \kappa_{pq}}
 - \ii\pdv{ }{\Im \kappa_{pq}})E(\bkappa)
 = 2(\lambda_{qp}-\lambda_{pq}^*),
\end{equation}

%\newpage
%\pagenumbering{gobble}

and the Hessian (see Appendix \ref{app:hessian} for its expression in the case of real orbitals) 
\begin{equation}
\label{eq:hessian}
\begin{split}
  G_{pq,rs} 
  & = {\eval{\pdv[2]{E(\bkappa)}{\kappa_{pq}^*}{\kappa_{rs}}}_{\bkappa=\bO}}
  \\
  & =\frac{1}{4}\qty(\pdv{ }{\Re \kappa_{pq}}+ \ii\pdv{ }{\Im \kappa_{pq}} ) \qty(\pdv{}{\Re \kappa_{rs}}- \ii\pdv{}{\Im \kappa_{rs}} ) E(\bkappa) 
\\
  & = \tG_{pq,rs}-\tG_{qp,rs}-\tG_{pq,sr}+\tG_{qp,sr} 
  -\tG_{pq,rs}-\tG_{qp,rs}-\tG_{pq,sr}-\tG_{qp,sr} 
  \\
  & + 2(\tG_{pq,sr}-\tG_{qp,rs}) 
  \\
  & = -4\tG_{qp,rs},
\end{split}
\end{equation}
with $\tG_{pq,rs}$ defined as 
\begin{equation}
\label{eq:Gpqrs_tilde}
\begin{split}
  \tG_{pq,rs} 
  & = \mel{\Psi}{\comm{\comm{\hH}{\sum_{\sigma} \cre{c}{p\sigma} \anh{c}{q\sigma}}}{\sum _{\sigma'} \cre{c}{r\sigma'} \anh{c}{s\sigma'}}}{\Psi} 
  \\
  & = \frac{1}{2} \qty[ \delta_{qr} (\lambda_{ps}+\lambda_{sp} ^*-4 n_r h_{sp})  +  \delta_{ps} (\lambda_{rq}+\lambda_{qr} ^*-4 n_p h_{qr} ) ] 
  \\
  & - 2 \qty[ {}^2D_{qr,qr}\braket{qs}{pr} +{}^2D_{rq,qr}\braket{sq}{pr} + {}^2D_{rr,qq}\qty( \braket{sr}{pq} + \braket{rs}{pq} ) ] (1-\delta_{qr} )
  \\
  & - 2 \qty[ {}^2D_{ps,ps}\braket{qs}{pr} +{}^2D_{ps,sp}\braket{qs}{rp}+ {}^2D_{pp,ss}\qty( \braket{qp}{rs} + \braket{qp}{sr} )](1-\delta_{ps}) 
  \\
  & + 2 \qty( {}^2D_{qs,qs}\braket{qs}{pr} +{}^2D_{sq,qs}\braket{sq}{pr}  ) (1-\delta_{qs}) 
  \\
  & + 2 \qty( {}^2D_{pr,pr}\braket{qs}{pr} +{}^2D_{pr,rp}\braket{qs}{rp}  ) (1-\delta_{pr}) 
  \\
  & + 2\delta_{qs} \sum_t {}^2D_{tt,qq} \braket{tt}{pr}+2\delta_{pr} \sum_t {}^2D_{pp,tt}\braket{qs}{tt} 
  \\
  &  -2\delta_{qr} \sum_t \qty( {}^2D_{rt,rt}\braket{st}{pt} +{}^2D_{tr,rt}\braket{ts}{pt} )
  \\
  & - 2\delta_{ps} \sum_t \qty({}^2D_{pt,pt}\braket{qt}{rt}+{}^2D_{pt,tp}\braket{qt}{tr} ),  
\end{split}
\end{equation}
in RDMFT and pCCD, we may approximate the energy by the following second-order Taylor series expansion
\begin{equation}
    E(\bkappa) 
    \approx E(\bkappa=0) 
    + {\bkappa}^\dag \cdot \bg +\frac{1}{2} \bkappa^\dag \cdot \bG \cdot \bkappa,  
\end{equation}
which is widely used in quadratic convergent methods and similar algorithms by updating the parameters $\kappa_{pq}$ with the Newton-Raphson step $\bkappa =-\bG^{-1} \cdot \bg$.\cite{helgaker:00book,dyall2007introduction,jensen1984direct,yamaguchi1990use,bozkaya2011quadratically,bozkaya2013orbital,boguslawski2014efficient,boguslawski2014nonvariational,boguslawski2014projected,henderson2014seniority,elayan2022deltano,nottoli2021second,cartier2024exploiting} At the stationary point, the gradient vector vanishes ($\bg = \bO$) and the diagonalization of the Hessian matrix $\bG$ provides valuable information about the type of stationary point one has reached: it is a minimum when all the eigenvalues are positive, a $k$th-order saddle point when there are $k$ negative eigenvalues, or a maximum when all eigenvalues are negative. Interestingly, this algorithm has notably been applied to optimize both occupation numbers and orbitals in RDMFT by (i) including the energy gradient with respect to the occupation numbers, and (ii) building an extended Hessian matrix, which incorporates the second derivative of the energy with respect to the occupation numbers, along with the corresponding crossed terms. The use of this algorithm is motivated by its accelerated convergence \cite{herbert:03jcp,cartier2024exploiting} at the expense of increasing the computational resources required for the storage and computation (see Sec.~\ref{sec:theo_res} for more details).   

\section{Theoretical consequences of incorporating time-reversal symmetry with complex orbitals in RDMFT and pCCD}
\label{sec:theo_res}

Enforcing time-reversal symmetry does not alter the energy contributions [see Eqs.~\eqref{eq:E_jkl} and \eqref{eq:E_jkl_pccd}] involving $J$ and $K$ integrals, but the energy contributions involving $L$ integrals in Eqs.~\eqref{eq:E_jkl} and~\eqref{eq:E_jkl_pccd} become contributions involving $K$ integrals. To show this, let us write the energy contribution involving $L$ integrals including the spin as
\begin{equation}
    \sum _{pq} {}^2D_{pp,qq} L_{pq}=  \sum _{pq}\sum_{\sigma,\sigma'=\uparrow,\downarrow \atop \sigma \neq \sigma'} {}^2D^{\sigma \sigma'}_{pp,qq} \iint \dd\br\dd\br' \frac{\chi_{p\sigma}^{*} (\br)\chi_{p\sigma'}^{*} (\br')\chi_{q\sigma} (\br)\chi_{q\sigma'} (\br') }{\abs{\br-\br'}},
\end{equation}
where the spin restriction $\sigma\neq\sigma'$ in the 2-RDM is a consequence of the Pauli exclusion principle. \rev{Then, let us write the two-electron repulsion integral for $\sigma=\uparrow$ and $\sigma'=\downarrow$ as two-component integrals
\begin{equation}
\begin{split}
& \iint \dd\br\dd\br' \frac{\chi_{p\uparrow}^{*} (\br)\chi_{p\downarrow}^{*} (\br') \chi_{q\uparrow} (\br)\chi_{q\downarrow}  (\br') }{\abs{\br-\br'}} \\
&= \iint \dd\br\dd\br' \frac{\left( \mqty( \chi_{p} (\br) \\ 0 ) \otimes  \mqty( 0 \\ \chi_{p} (\br') ) \right)^\dagger \cdot \left( \mqty( \chi_{q} (\br) \\ 0 ) \otimes  \mqty( 0 \\ \chi_{q} (\br') ) \right) }{\abs{\br-\br'}} \\
&=  \iint \dd\br\dd\br' \frac{\left( \mqty( \chi_{p} (\br) \\ 0 ) \otimes \hTheta \mqty( \chi_{p}^{*}  (\br') \\ 0) \right)^\dagger \cdot \left( \mqty( \chi_{q} (\br) \\ 0 ) \otimes \hTheta \mqty( \chi_{q}^{*} (\br') \\ 0 ) \right) }{\abs{\br-\br'}} \\
&=  \iint \dd\br\dd\br' \frac{\left( \left(\mathcal{I}_2 \otimes \hTheta\right) \cdot \mqty( \chi_{p} (\br) \\ 0 ) \otimes  \mqty( \chi_{p}^{*}  (\br') \\ 0) \right)^\dagger \cdot \left( \left(\mathcal{I}_2 \otimes \hTheta\right) \cdot \mqty( \chi_{q} (\br) \\ 0 ) \otimes \mqty( \chi_{q}^{*} (\br') \\ 0 ) \right) }{\abs{\br-\br'}} \\
&=  \iint \dd\br\dd\br' \frac{\left( \mqty( \chi_{p} (\br) \\ 0 ) \otimes  \mqty( \chi_{p}^{*}  (\br') \\ 0) \right)^\dagger \cdot \left(\mathcal{I}_2 \otimes \hTheta\right)^\dagger \cdot \left(\mathcal{I}_2 \otimes \hTheta\right) \cdot \left(  \mqty( \chi_{q} (\br) \\ 0 ) \otimes \mqty( \chi_{q}^{*} (\br') \\ 0 ) \right) }{\abs{\br-\br'}} \\
  &= \iint \dd\br\dd\br' \frac{\chi_{p\uparrow}^{*} (\br)\chi_{q\uparrow}^{*} (\br')\chi_{q\uparrow} (\br) \chi_{p\uparrow} (\br') }{\abs{\br-\br'}},
 \end{split}
\end{equation}
where we used Eq.~\eqref{eq:time_rev_def}, $\otimes$ is the tensor product,  $\dagger$ implies complex conjugation and transposition, $\mathcal{I}_2$ is the 2$\times 2$ identity matrix, and we relied on the fact that $\hTheta^\dagger \cdot \hTheta = \mathcal{I}_2$ to obtain the final expression (after some trivial reorganization).} Therefore, the first consequence of imposing time-reversal symmetry is that the energy contributions involving the $L$ integrals become contributions involving (real) $K$ integrals
\begin{equation}
  \sum _{pq} {}^2D_{pp,qq} L_{pq} =\sum _{pq} {}^2D_{pp,qq} K_{pq},
\end{equation}
which introduces a simplification of the RDMFT and pCCD energy expression that can be written as a $JK$-only functional (as in the real case).

Next, let us focus on the $t$- and $z$-amplitude equations of the pCCD method [see Eqs.~\eqref{eq:t_amp} and \eqref{eq:z_amp}]. In both equations, $L$ integrals are present (and involve interactions among opposite-spin electrons). Hence, adopting time-reversal symmetry, we may replace $L$ integrals with $K$ integrals making the $t$ and $z$ amplitudes real-valued also for complex orbitals and even for complex 2-spinors that are related via time-reversal. The latter consequence can be derived by approximating the Kramers-restricted CCSD formalism\cite{visscher1995KRCCSD} to pCCD. By taking only paired excitations only one of the three excitation classes survives and these amplitudes are real because $(t_{i\bar{i}}^{a\bar{a}})^* = \hTheta t_{i\bar{i}}^{a\bar{a}}= t_{\bar{i}i}^{\bar{a}a} = t_{i\bar{i}}^{a\bar{a}}$, where we have labeled as barred and unbarred the 2-spinors related by time-reversal symmetry.  Consequently, the 1RDM and 2RDM elements also become real. However, the Hermiticity of the 2RDM elements is not guaranteed (i.e., ${}^2D_{pp,qq}\neq{}^2D_{qq,pp}$) because left- and right-hand wave functions, $\bra{\mathcal{L}} e^{-\hT}$ and $e^{\hT} \ket{0}$, respectively, are used to build these elements. Nevertheless, numerical evidence indicates that the deviation from Hermiticity of the 2RDM is usually small \cite{henderson2014seniority}. In addition, one can always impose the Hermiticy of these elements by averaging the elements ${}^2D_{pp,qq}$ and ${}^2D_{qq,pp}$ before entering the orbital optimization process. The value of the energy is not affected by this averaging because of the replacement of the $L$ integrals by real-valued $K$ integrals that are symmetric with respect to index exchange (i.e. $K_{pq} = K_{qp}$). Furthermore, imposing the Hermiticity of the 2RDM elements makes Eq.~\eqref{eq:lpp_lpp} equal to zero also for the pCCD approximation, which is a crucial condition for using the optimization procedure presented in Sec.~\ref{sec:orb_opt}.

To analyze the next consequence, let us focus on the diagonal terms of the gradient $g_{pp}=2(\lambda_{pp}-\lambda^*_{pp})$, as given by Eq.~\eqref{eq:gradient_elements}. These are iteratively reduced towards zero thanks to an orbital phase adjustment originating from the optimization parameters $\ii\Im \kappa_{pp}$. This can be illustrated by considering the particular case where both the gradient and Hessian matrices have a diagonal structure, that is, $g_{pp} \neq 0$ and ${G}_{pq,pq}\neq 0$. In this case, the unitary matrix $e^{\bkappa}$ is constructed using a matrix $\bkappa$ that only contains the diagonal elements $\ii\Im \kappa_{pp}$. This only alters the orbital phases during the self-consistent procedure, i.e., $\chi_p(\br) \leftarrow \chi_p(\br) e^{-\ii \Im \kappa_{pp}}$. When time-reversal symmetry is not enforced, we have $g_{pp}\neq 0$ which involves that the phases of the orbitals must be optimized because the diagonal elements of the gradient must be, by definition, zero at the stationary solution. On the contrary, imposing time-reversal symmetry yields $g_{pp} = 0$ when the real-valued $K$ integrals replace the $L$ integrals. 

Next, let us focus on the off-diagonal elements $g_{pq}$ ($p \neq q$), which are the ``active gradients'' that must vanish at the stationary solution when one imposes time-reversal symmetry. The ``active gradients'' are related to the off-diagonal elements of the matrix $\bF$ defined in the Piris-Ugalde algorithm as $F_{pq} = g_{pq}/2$. \cite{burton2020energy,douady1980exponential} For this reason, the Piris-Ugalde algorithm can also be employed to optimize the complex-valued orbitals in RDMFT and pCCD methods when time-reversal symmetry is imposed. Furthermore, in this algorithm, the diagonalization of $\bF$ (see Fig.~\ref{fig:scf_algo}) produces a unitary matrix $\boldsymbol{U}$ that transforms the natural orbital coefficients $\bc$ from one iteration to the other as $\bc \leftarrow \boldsymbol{U} \cdot \bc$, making the gradient elements equal to zero (i.e., $g_{pq}=0$ for $p \neq q$) in this direction (and iteration). Consequently, the Piris-Ugalde algorithm is equivalent to a gradient-descent method, which explains the large number of iterations observed near the stationary solutions when compared to quadratic convergent methods \cite{jensen1984direct,jensen1986direct,helgaker1986molecular} or methods that use an approximate Hessian matrix. \cite{kreplin2020mcscf,cartier2024exploiting,vidal2024geometric} 

\rev{It is worth mentioning that a large number of iterations is observed only close to the solution. For example, for \ce{N2} computed with the cc-pVDZ basis set~\cite{dunning:89jcp} at $R_{\textrm{N-N}}=0.7$ \AA, the final converged GNOF energy ($E=-107.688637$ a.u.) is obtained after 276 occupation number optimizations and 8228 orbital rotations (when using real orbitals). Although, after 102 occupation number optimizations and 3060 orbital rotations, the energy is $E=-107.688093$ a.u., differing by approximately 5$\times 10^{-4}$ a.u.~from the final converged value. Hence, most of the optimization steps are due to the slow convergence close to the solution. On the other hand, when non-dynamic electronic correlation effects are dominant, the optimization is much faster. For example, for \ce{N2} at $R_{\textrm{N-N}}=6.0$ \AA, the occupation numbers of the broken bonds rapidly adjust to their limiting value of $0.5$, changing less during each iteration compared to regions dominated by dynamic correlation effects. In this case, convergence is achieved after 21 occupation number optimizations and 150 orbital rotations, starting from orbitals obtained by diagonalizing the core Hamiltonian. Finally, let us highlight that novel, better, and faster optimization algorithms \cite{cartier2024exploiting,huanlewyee2025:adam_opt,vidal2024geometric} are rapidly spreading in the RDMFT community, allowing the computation of large systems and pushing the limits of the applicability of this theory, without needing to build explicitly the Hessian matrix used in quadratic convergent methods.}

In terms of computational cost, the construction of $\bF$ in the Piris-Ugalde algorithm scales as $M^4$ and requires $M^2$ storage when density fitting approximations are employed. \cite{lew2021resolution} (The bottleneck here is the transformation of the electron repulsion integrals to the orbital basis.) On the contrary, quadratic convergent methods require the computation of the Hessian matrix. For the exact Hessian, the computational cost associated with its construction scales as $M^5$ [see Eq.~\ref{eq:Gpqrs_tilde}] and $M^4$ for its storage, which makes it prohibitively expensive for large systems where the Piris-Ugalde algorithm should be preferred. Note that the computational cost can be lowered to $M^4$ at the expense of defining additional intermediates that would further increase storage.

As a summary of the theoretical consequences, incorporating time-reversal symmetry in complex orbitals within RDMFT and pCCD implies that $L$ integrals can be replaced by $K$ integrals, which permits us to employ the Piris-Ugalde algorithm to perform the orbital optimization. \cite{piris2009iterative} Moreover, for pCCD calculations, the $\bt$ and $\bz$ amplitudes become real-valued quantities due to the replacement of $L$ integrals by real-valued $K$ integrals in the amplitude equations.  

\section{Numerical consequences of incorporating time-reversal symmetry with complex orbitals in RDMFT and pCCD}
\label{sec:prac_res}

To analyze the practical consequences, we present some calculations performed with representative systems at different geometries, which lead to different flavors of electronic correlation effects.  

\subsection{Computational details}

All calculations presented in this work were performed with the \texttt{MOLGW} program \cite{bruneval2016molgw} that incorporates the stand-alone NOFT module \cite{rodriguez2022snoft} based on the \texttt{DoNOFT} program \cite{piris2021donof} that performs RDMFT and pCCD calculations \rev{with the Piris-Ugalde algorithm}. \cite{piris2009iterative} For this study, we have incorporated the pCCD method and the use of complex orbitals including time-reversal symmetry into the NOFT module. The calculations on the \ce{H2}, \ce{LiH}, \ce{BH}, and \ce{N2} molecules were performed using the cc-pVDZ basis set \cite{dunning:89jcp} including density fitting techniques. \rev{The FCI energies for \ce{N2} were taken form Ref.~\onlinecite{eriksen2019many}.} For the \ce{BeH2} system, the basis set developed by Evangelista and collaborators \cite{evangelista2010insights} was employed to facilitate the comparison with previous studies. \cite{ammar2024can,mitxelena2024assessing} We have labeled the real restricted solutions as RHF, RPNOF5~\cite{piris:11jcp}, RPNOF7~\cite{piris2018dynamic}, RGNOF~\cite{piris:21gnof}, and RpCCD, while for the complex solutions including time-reversal symmetry, we denote them as THF, TPNOF7, TPNOF5, TGNOF, and TpCCD. The THF results correspond to the axial spin current wave in Fukutome's labeling \cite{fukutome1981unrestricted} or paired unrestricted HF in Stuber-Paldus designation. \cite{stuber2003symmetry}

We employed either the real orbitals produced with the Perdew–Burke–Ernzerhof density-functional approximation \cite{perdew:prl96} or the diagonalization of the core Hamiltonian as the starting point for the RDMFT and pCCD calculations. For calculations using complex-valued orbitals, the real orbitals were multiplied by random imaginary phases $e^{\ii\theta}$ with $\theta \in [0,2\pi)$ being a random number. All electrons were included in the active space except for \ce{N2}, where the $1s$ electrons were frozen. Also, all virtual orbitals were included in the active space. Finally, let us highlight that, for each studied system and method, we have evaluated the eigenvalues of the real and complex Hessian matrices in the regions where the real and the complex solutions differ (see below) to confirm that the targeted solution corresponds to a minimum. We have also tested different starting points to make sure that the solution found was the global minimum. 

\rev{Finally, let us mention that a comprehensive benchmark of complex-valued RDMFT methods including larger systems and the most recently developed optimization algorithm~\cite{huanlewyee2025:adam_opt} is left for future work.}

\subsection{When the complex (with time-reversal symmetry) and the real solutions coincide.}
\label{sec:h2_lih}

\begin{figure}[H]
\centering\includegraphics[scale=0.8]{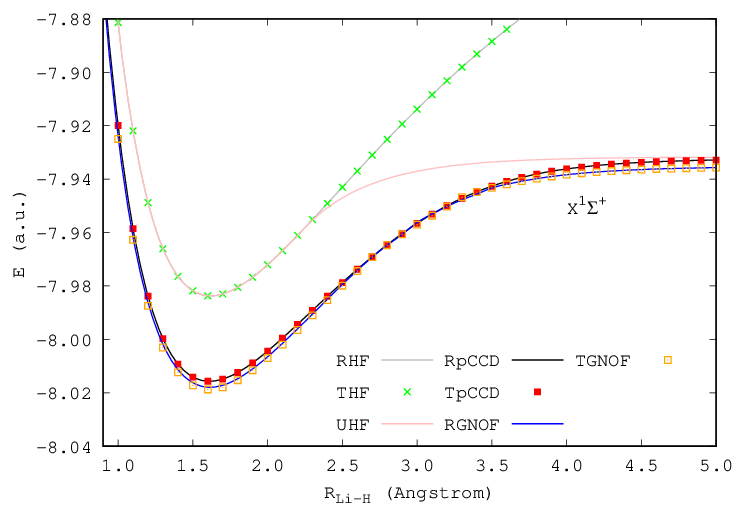}
    \caption{Potential energy curves obtained with the real (solid) and complex (dotted) versions of the HF, pCCD, and GNOF methods for the dissociation of the \ce{LiH} molecule.} 
    \label{fig:lih_pec}
\end{figure}

For some systems (e.g., \ce{H2} and \ce{LiH}), the use of complex orbitals does not provide any extra flexibility and the restricted real solutions coincide with the time-reversal-symmetric complex orbitals. Here we focus on the \ce{LiH} case. (See the supplementary material for the \ce{H2} example.) In Fig.~\ref{fig:lih_pec}, we have represented the potential energy curve (PEC) for the homolytic dissociation of \ce{LiH} obtained with HF, pCCD, and the GNOF RDMFT functional approximation. The real unrestricted HF (UHF) PEC is also included for comparison purposes. In \ce{LiH}, only one pair of electrons forms the bond while the $1s^2$ electrons of \ce{Li} remain almost unaltered at all bond lengths. In the ground state, the bond is formed by the so-called harpoon mechanism, \cite{rodriguez:16mp} where the dominant species are \ce{Li}$^+$ and \ce{H}$^-$ around the equilibrium distance while neutral atoms are formed in the dissociation limit. As we can observe, the GNOF functional and the pCCD approximation results are similar because both methods accurately describe the correlation effects of the electron pair responsible for forming the bond. In addition, as shown in Fig.~\ref{fig:lih_pec}, the real solutions coincide with the complex ones for all bond lengths along the dissociation curve. The analysis of the eigenvalues of the complex Hessian matrix revealed that the real orbitals also lead to a minimum for the complex orbital optimization problem. Also, the analysis of the electronic density shows that the real and the complex electronic densities coincide with only tiny numerical differences caused by the finite convergence thresholds. 

\subsection{When the complex (with time-reversal symmetry) and the real solutions differ.}
\label{sec:beh2_n2}

\begin{figure}[H]
\begin{subfigure}[b]{0.45\textwidth}
\includegraphics[scale=1.0]{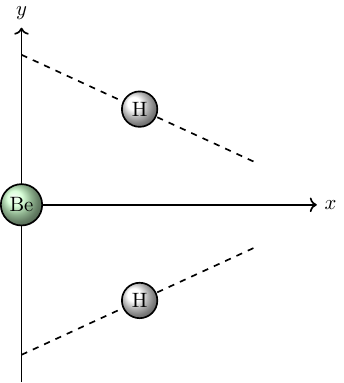}
    \caption{}
    \label{fig:sketch_beh2}    
\end{subfigure}
\hfill
\begin{subfigure}[b]{0.45\textwidth}
\includegraphics[scale=0.7]{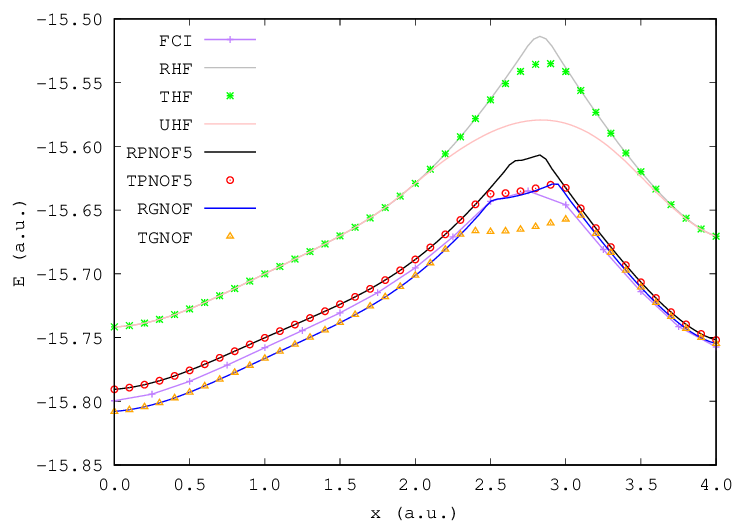}
    \caption{}
    \label{fig:beh2_pec} 
\end{subfigure}
\begin{subfigure}[b]{0.45\textwidth}
\includegraphics[scale=0.7]{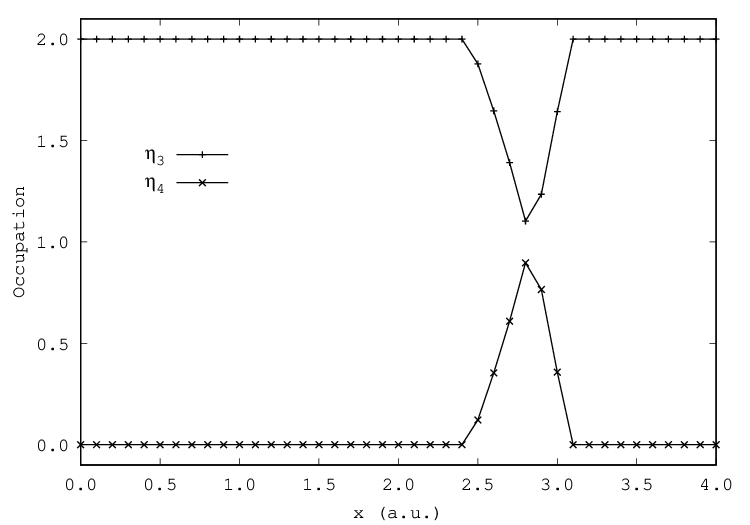}
    \caption{}
    \label{fig:beh2_rho}    
\end{subfigure}
\hfill
\begin{subfigure}[b]{0.45\textwidth}
\includegraphics[scale=0.7]{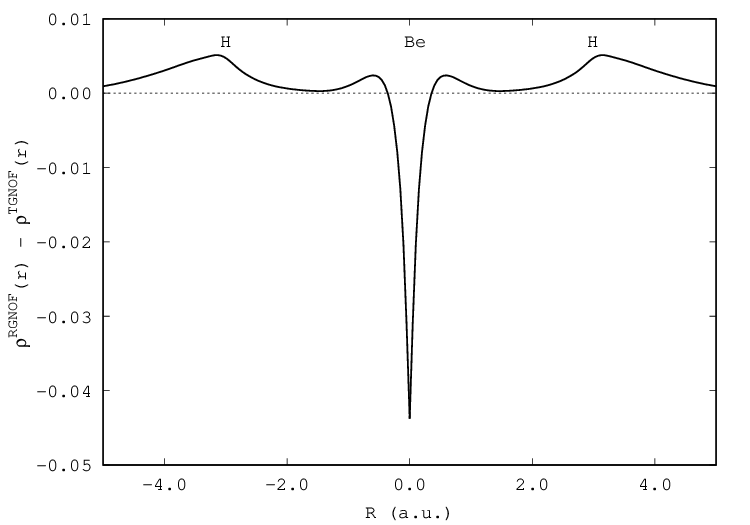}
    \caption{}
    \label{fig:beh2_occ} 
\end{subfigure}
    \caption{\small{(a) Schematic representation of the insertion reaction of \ce{Be} into \ce{H2} to form \ce{BeH2}. The \ce{Be} atom is placed at the origin and the hydrogen atoms are located at $\pm y = 2.54 - 0.46x$. (b) Potential energy curves obtained with the real (solid) and complex (dotted) versions of HF, PNOF5, and GNOF methods for $ 0 \le x \le 4$. The real UHF and the FCI curves are also included for comparison. (c) Changes in the spin-summed occupations numbers ($\eta_3$ and $\eta_4$) for the THF natural orbitals as functions of the reaction coordinate $x$. (d) Difference between the real and complex optimized electronic densities, $\rho(\br)$, for the GNOF functional approximation along the insertion pathway for $x=2.75$ Bohr.}}
\end{figure}

While the time-reversal-symmetric complex and real solutions match for the \ce{H2} and the \ce{LiH} systems across all bond lengths, this correspondence does not necessarily occur in other systems. To illustrate this, we have studied the PEC of \ce{BeH2} during the insertion of a beryllium atom into a hydrogen molecule. In Fig.~\ref{fig:sketch_beh2}, we have represented the reaction coordinate $x$ (in Bohr), where the \ce{Be} atom is placed at the coordinate origin and the H atoms are located at $\pm y = 2.54 - 0.46x$ with $x \in [0,4]$. This system has recently been used as a benchmark tool of different methods~\cite{boguslawski2014nonvariational,ammar2024can,gaikwad2024coupled,mitxelena2024assessing,gdanitz1988averaged,mahapatra1998state,mahapatra1999size,sharp2000sigma,kallay2002general,pittner2004performance,ruttink2005multireference,lyakh2006multireference,yanai2006canonical,evangelista2010insights,evangelista2011alternative,evangelista2011orbital} including the pCCD and RDMFT functional approximations, where the ability of these methods to account for non-dynamic electronic correlation effects was evaluated. For small values of $x > 0$, the exact wave function is primarily governed by the electronic configuration \(\ket{(1a_1)^2 (2a_1)^2 (1b_2)^2}\). As $x$ increases, however, the configuration \(\ket{(1a_1)^2 (2a_1)^2 (3a_1)^2}\) becomes dominant. 
In the range $2.5 < x < 3$, the wave function undergoes a rapid transition from \(\ket{(1a_1)^2 (2a_1)^2 (1b_2)^2}\) to \(\ket{(1a_1)^2 (2a_1)^2 (3a_1)^2}\).
Therefore, in the \ce{BeH2} system, the region $ 2.5 \le x \le 3.5$ exhibits strong non-dynamic electronic correlation effects while the dynamic component is dominant for all other geometries.

Focusing on the consequences of using complex orbitals with time-reversal symmetry, we see in Fig.~\ref{fig:beh2_pec} that in regions where the dynamic electronic correlation effects are dominant the real and the complex solutions coincide. On the contrary, the flexibility provided by the complex natural orbital coefficients leads to a relaxation of the electronic density in the region where non-dynamic correlation effects are dominant (i.e., $x \in \left[2.5,3.5\right]$) making the complex solutions lie below the real ones for all methods studied. Let us first analyze the HF solutions. As we can observe, both solutions produce a smooth curve in the region where non-dynamic electronic correlation effects are dominant with the THF solution lying below the RHF one only on a very small interval. Using the real RHF orbitals to build the complex Hessian matrix [see Eq.~\eqref{eq:hessian}] in the interval where the solutions differ and proceeding to diagonalize it, we obtain one or two (depending on the geometry) negative eigenvalues. Since the gradient is still zero, this result indicates that the real solution is a stationary point (i.e., a saddle point) for the complex optimization problem. Thus, by re-optimizing the orbitals (and occupation numbers) we obtain the actual minimum. This result is comparable to the usual ones obtained with restricted and unrestricted methods for geometries beyond the Coulson-Fisher point. \cite{burton2020holomorphic} However, as we show in Fig.~\ref{fig:beh2_pec}, the THF energy lies above the real UHF one, which shows that the flexibility provided by the complex orbital coefficients is not sufficient to account for all the non-dynamic electronic correlation effects present. 

To gain more insights into the THF solution, we have computed the spin-summed occupations numbers for the THF natural orbitals as a function of the reaction coordinate $x$. Note that this procedure is equivalent to the construction of the spin-summed unrestricted natural orbitals within the unrestricted HF formalism. \cite{helgaker:00book} To do so, we built \rev{
\begin{equation}
   \boldsymbol{P} = \bc_{\uparrow} ^\dagger \cdot {}^1\boldsymbol{D} \cdot \bc_{\uparrow} + \bc_{\downarrow}^\dagger \cdot {}^1\boldsymbol{D} \cdot \bc_{\downarrow} 
\end{equation} }
where $\boldsymbol{P}$ is the density matrix written in the real (scalar) atomic orbital basis $\psi$, $\boldsymbol{c}_{\uparrow}$ ($\boldsymbol{c}_\downarrow$) gathers the molecular orbital coefficients for the spin-up (spin-down) orbitals, and ${}^1\boldsymbol{D}$ is the HF first-order reduced density matrix (with ${}^1{ D}_{pq} = \delta_{pq}$ for $p,q$ occupied and 0 otherwise). Then, using the L\"owdin orthonormalization ($\boldsymbol{S}^{-1/2} \cdot \boldsymbol{P} \cdot  \boldsymbol{S}^{-1/2} $ with $\boldsymbol{S}$ being the overlap matrix of the real (scalar) orbitals) and diagonalizing the resulting matrix, we obtained the THF natural orbitals and THF natural occupation numbers (${\boldsymbol{\eta}}$). In Fig.~\ref{fig:beh2_occ}, we have plotted the spin-summed occupations numbers for the THF natural orbitals for the 3rd and 4th natural orbitals ($\eta_3$ and $\eta_4$). The $\eta_1$ and $\eta_2$ occupation numbers remain equal to 2 for all geometries. As we can observe in Fig.~\ref{fig:beh2_occ} in the region where non-dynamic correlation effects are dominant, $\eta_3$ and $\eta_4$ approach 1. Therefore, the THF is capable of retrieving some non-dynamic correlation effects present when compared to RHF, but its ability is limited and it is unable to perform as well as real UHF in this region (see Fig.~\ref{fig:beh2_pec}).

Moving to the PNOF5 and GNOF results, we also observe that the real and the complex solutions differ in the region where non-dynamic correlation effects are dominant. In the case of PNOF5, which is a fully $N$-representable method \cite{garrod:64jmp,mazziotti2016pure} thanks to its correspondence with the constrained anti-symmetrized product of strongly orthogonal geminals, \cite{pernal:13ctc} we notice that the TPNOF5 estimates (red dots) lie above the FCI energies. \rev{Interestingly, the PNOF5 results are very similar to the pCCD ones due to the close relationship between their associated wavefunctions. For this reason, the pCCD results for this system are reported and discussed in the supplementary material.} In the case of the TGNOF results (orange dots), these lie below their FCI counterparts in the $x \in \left[2.5;3.5\right]$ interval, which indicates that this functional approximation can introduce $N$-representability violations when non-dynamic correlation effects are pronounced.  Next, the analysis of the difference in the electronic density along the \ce{H-Be-H} path for $x=2.75$ Bohr reveals that, for the GNOF functional, non-negligible changes occur in the vicinity of the nuclei and in the bonding region. (Similar results were obtained with the PNOF5 functional approximation.) Finally, let us mention that contrary to the HF case, the analysis of the eigenvalues of the complex Hessian matrix built with the RPNOF5 and RGNOF occupation numbers and orbitals revealed that the real solutions correspond to local minima for the complex optimization problem because all the eigenvalues obtained were positive.    

To gain further insights into the complex solutions, let us analyze the THF as the non-relativistic limit of the Kramers' restricted four-component DIRAC-HF (KR-4c-DHF) equation. \cite{dyall2007introduction,saue2011relativistic,hafner1980kramers} The KR-4c-DHF equation is the relativistic extension of the HF method for relativistic calculations, which produces 4c-spinors preserving time-reversal symmetry. It is known that the non-relativistic limit can be approached by setting the value of the speed of light $c$ to a large value in the 4c-DHF. \cite{dyall2007introduction,saue2011relativistic,pyykko2012physics,rodriguez2022relatintrac} Then, as discussed in Sec.~\ref{sec:theo}, in this limit the KR-4c-DHF solution approaches the THF solution (instead of the RHF one) when the RHF and THF solutions differ because (i) the THF solution can be lower in energy, (ii) the THF and the KR-4c-DHF methods work with the extra flexibility provided by complex orbitals, and (iii) both methods are built to preserve time-reversal symmetry. To show this, we have taken the \ce{BeH2} system at the $x=2.75$ Bohr geometry, employed the cc-pVDZ basis set, \cite{dunning:89jcp} and performed calculations using the \texttt{DIRAC} program \cite{saue2020dirac} setting the speed of light value to $c=10^5$. In the non-relativistic limit, the KR-4c-DHF energy ($-15.57555$ hartree) approaches the THF value ($-15.575600$ hartree), where for $c=10^5$ the energy difference is lower than $5 \times 10^{-5}$ hartree (The RHF energy is $-15.563664$ hartree). This result illustrates that KR-4c-DHF solutions in the non-relativistic limit only approach the RHF ones when the RHF and the THF solutions are equivalent (which occurs in regions where non-dynamic correlation effects are not dominant). Otherwise, the KR-4c-DHF solutions in the non-relativistic limit may recover the THF values. \bibnote{Similar results were obtained for lowest singlet state of \ce{O2} at a distance of $R_{\text{O-O}}=10$ {\AA}, where the RHF energy is $-149.33007$ hartree, the THF energy is $-149.41117$ hartree, and the KR-4c-DHF($c=10^7$) energy is $-149.41072$ hartree with the difference between the THF and the KR-4c-DHF being $4.5\times10^{-4}$ hartree.}  

\begin{figure}[H]
\includegraphics[scale=1.0]{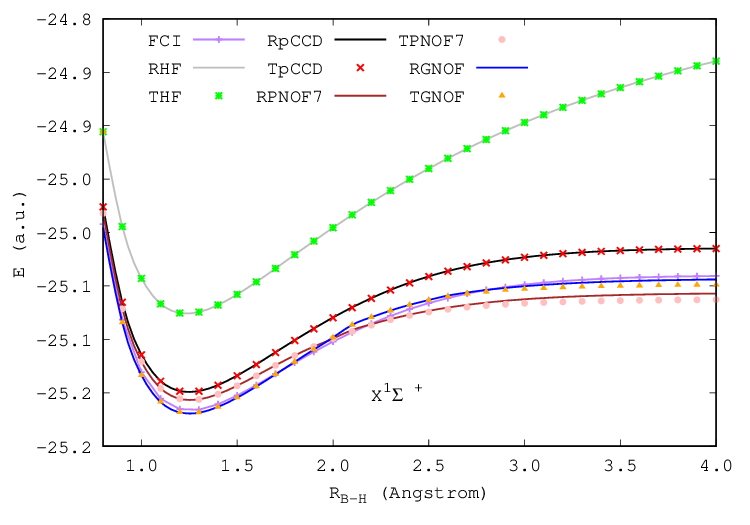}
    \centering
    \caption{Potential energy curves obtained with the real (solid) and complex (dotted) versions of the HF, pCCD, PNOF7, and GNOF methods for the dissociation of the \ce{BH} molecule.}
    \label{fig:bh_pecs}    
\end{figure}

\rev{Another example where the real and the complex solutions may differ is the \ce{BH}  homolytic dissociation. In Fig.~\ref{fig:bh_pecs} we have collected the potential energy curves obtained with real-valued and complex-valued orbitals using the HF, pCCD, PNOF7, and GNOF methods. As in our previous example, the pCCD and the PNOF5 results are very similar; thus, we have only included the pCCD values. As we can observe from Fig.~\ref{fig:bh_pecs}, only for the PNOF7 and GNOF approximations the use of complex-valued orbitals leads to different solutions in the region where non-dynamic electronic correlation effects are dominant. On the contrary, for the HF, pCCD, and PNOF5 methods, both solutions coincide. Compared to the FCI results, the potential energy curve of the RGNOF approximation is the most accurate one; then, the performance of the TGNOF deteriorates when the \ce{BH} bond is broken because TGNOF produces lower energies in this region. Note that the TGNOF, RPNOF7, and TPNOF7 energies lie below the FCI values, which shows that these functional approximations introduce $N$-representability violations. Let us recall that for the \ce{BH} system it is possible to find a generalized HF potential energy curve~\cite{lowdin1992some} that differs from the RHF one. Nevertheless, as shown in Fig.~\ref{fig:bh_pecs}, the THF method does not have enough flexibility (compared to the generalized HF one); therefore, the potential energy curve obtained with the THF method coincides with the RHF one. Recalling that \ce{BeH2} and \ce{BH} present generalized HF solutions, and that for these systems we obtained different solutions using the RDMFT functionals and/or the pCCD method, we may argue that there is a weak relationship between the presence of a generalized HF solution and the discovery of a new solution when employing complex-valued orbitals in the RDMFT and pCCD methods. However, this relationship is not strong, as not all methods are consistently affected. For instance, the pCCD method does not yield a new solution for the \ce{BH} system when using complex-valued orbitals.}

\begin{figure}[H]
\begin{subfigure}[b]{0.45\textwidth}
\includegraphics[scale=0.7]{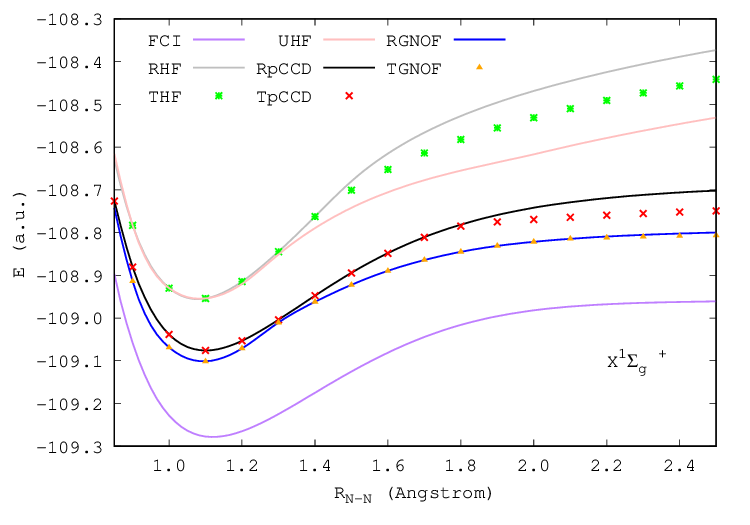}
    \caption{}
    \label{fig:n2_pecs}    
\end{subfigure}
\hfill
\begin{subfigure}[b]{0.45\textwidth}
\includegraphics[scale=0.7]{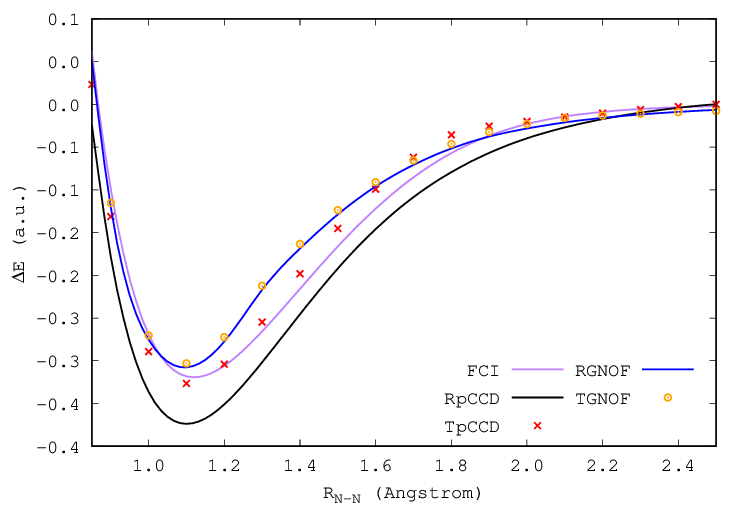}
    \caption{}
    \label{fig:n2_DE_pecs} 
\end{subfigure}
\caption{(a) Potential energy curves obtained with the real (solid) and complex (dotted) versions of the HF, pCCD, and GNOF methods for the dissociation of the \ce{N2} molecule. The real UHF results are included for comparison. The FCI values were taken from Ref.~\onlinecite{eriksen2019many}. (b) $\Delta E (\textrm{in \; a.u.}) = E_{\textrm{R}_{\textrm{N-N}}}-E_{\textrm{R}_{\textrm{N-N}}=2.5 \textrm{\AA}}$.}
\label{fig:n2_pec}
\end{figure}

Next, let us discuss our last example where the real and complex solutions differ, the homolytic dissociation of the \ce{N2} molecule in its ground state, where three pairs of electrons are simultaneously broken. In Fig.~\ref{fig:n2_pecs} we have collected the real and complex HF, pCCD, and GNOF PECs. As one can observe, for all methods the real and complex solutions are equivalent up to $R_{\text{N-N}}\sim 1.4$ \AA. Then, for larger interatomic distances, the real and the complex solutions differ, with the complex solution lying below the real one in all cases. As in the \ce{BeH2} example, the RHF is a saddle point of the complex solution when we evaluate the Hessian matrix. Again, the THF results can partially retrieve some non-dynamic electronic correlation effects but are still far from the UHF results. On the other hand, the GNOF real and complex results are very similar but the pCCD ones present large deviations, which shows that the difference between the real and the complex solutions is system- and method-dependent. Once more, as for the \ce{BeH2} system, all the eigenvalues of the complex Hessian matrix built with the real GNOF/pCCD occupation numbers/amplitudes and orbitals (multiplied by some random phases) were positive, which suggests that the real solutions correspond to local minima of the complex optimization problem for the RDMFT functional approximations and the pCCD method. \rev{Comparing the GNOF and pCCD methods against FCI (see Fig.~\ref{fig:n2_pecs}), we observe that real and complex results lie far from the FCI values because of the missing dynamic electronic correlation effects. Finally, concerning the shape of the potential energy curves, we notice that the TpCCD curve improves over the RpCCD curve for all geometries (see Fig.~\ref{fig:n2_DE_pecs}) and also improves for the value of the dissociation energy.}

\begin{figure}[H]
\includegraphics[scale=1.0]{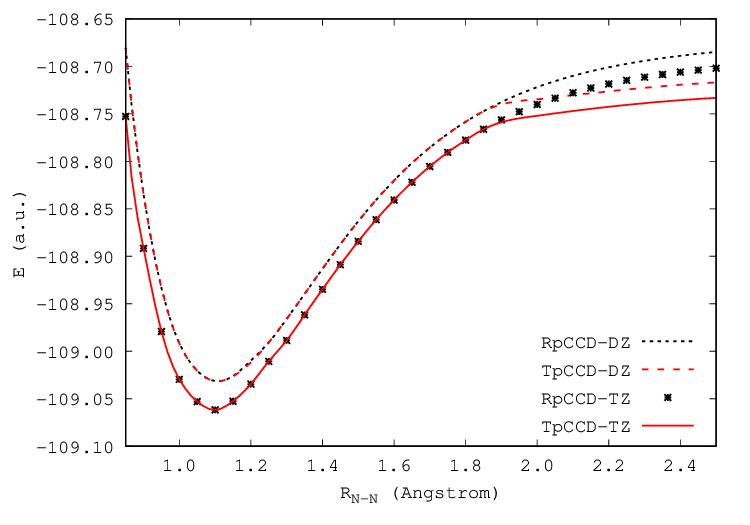}
    \centering
    \caption{Potential energy curves computed with the cc-pVDZ and cc-pVTZ basis sets and using the real (black) and complex (red) versions of the pCCD method for the homolytic dissociation of the \ce{N2} molecule.}
    \label{fig:n2_basis}    
\end{figure}

\rev{Up to now, we have shown the effect of using complex-valued orbitals with small basis sets for all systems to facilitate the analysis. Nevertheless, it is worth evaluating the role of the size of the basis set. To this end, we have collected in Fig.~\ref{fig:n2_basis} the potential energy curves for the \ce{N2} system computed using the cc-pVDZ and the cc-pVTZ basis sets, \cite{dunning:89jcp} and using the pCCD method. We only report the pCCD results because it leads to the largest difference between the real-valued and complex-valued orbitals solutions among all the methods employed. (For example, using the RGNOF and the TGNOF approximations the difference among the energies for the ${\textrm{R}}_{\textrm{N-N}}=6.0$ \AA~geometry is only $\sim 0.009$ a.u.). For this test, we have frozen the 1$s$ and 2$s$ electrons because the occupation numbers of these orbitals remain practically constant for all geometries. As we can observe, the RpCCD potential energy curve computed with the cc-pVDZ basis set is almost parallel to the cc-pVTZ one; the same holds for the TpCCD curves which indicates that the consequences of using complex-valued orbitals are consistent across different basis sets. Nevertheless, as we discuss in the following section, the use of complex-valued orbitals plays a crucial role in the convergence.}

\subsection{\rev{Changes on the convergence caused by using complex-valued orbitals.}}
\label{sec:conver_complx}
\rev{First of all, let us focus on the speed of convergence, which is affected by the random phases employed. For example, for \ce{N2} at $R_{\textrm{N-N}}=6.0$ \AA, taking as a starting point the real orbitals obtained from the diagonalization of the core Hamiltonian multiplied by random phases. The TGNOF can converge for two sets of random phases after (a) 21 occupation number optimizations and 305 orbital rotations, or (b) 36 occupation number optimizations and 501 orbital rotations. Consequently, the random phases in the non-dynamic electronic correlation regime strongly affect the speed of convergence. On the contrary, when dynamic electronic correlation effects are dominant, convergence is less affected by the random phases; a similar number of iterations is obtained with real- or complex-valued orbitals. In summary, the random phases employed to build the complex-valued orbitals determine the speed of converge. Similar results were obtained for other RDMFT approximations and the pCCD method.}  

\rev{Despite the random-phase dependency observed in the speed of convergence when using complex-valued orbitals, let us mention that convergence is facilitated by using complex-valued orbitals, especially, when non-dynamic electronic correlation effects are enhanced. For example, for the dissociated \ce{N2} system (i.e.~ $R_{\textrm{N-N}}>2.5$ \AA), convergence to the lowest energy state using pCCD and real orbitals was not possible with the algorithm employed (for the cc-pVDZ and the cc-pVTZ basis sets). Attempting to reach convergence, we used the PNOF5 solutions as a starting point for the pCCD calculations. Unfortunately, this strategy fails to converge when using real-valued orbitals. However, it is effective when complex-valued orbitals are employed. For instance, using the TPNOF5 orbitals as a starting point, we have converged the TpCCD energy after 38 amplitude optimizations and 1036 orbital rotations for $R_{\textrm{N-N}}=10$ \AA. Thus, in general, we have observed that working with complex-valued orbitals facilitates convergence when non-dynamic correlation effects become dominant.}

\section{Conclusions}
\label{sec:ccl}

In this work, we have presented and discussed the consequences of using complex orbitals including time-reversal symmetry in RDMFT and pCCD calculations. From a theoretical perspective, the RDMFT $JKL$-only functional approximations and the pCCD method reduce to $JK$-only methods where only the Hartree and exchange integrals are needed to evaluate the energy. Specifically, for spin-compensated systems, the energy expression is given by
\begin{equation}
    E^{\text{TPNOFs/TGNOF/TpCCD}} = 2 \sum _p n_p h_{pp} + \sum_{pq} \left( {}^2D_{pq,pq} J_{pq} +^2D_{pq,qp} K_{pq} \right). 
    \label{eq:Etrsym}
\end{equation}
This simplification occurs because the $L$ integrals that accompany opposite-spin interactions become $K$ integrals when time-reversal symmetry is considered. Consequently, the $t$ and $z$ amplitudes of TpCCD are also real-valued. Note that Eq.~\eqref{eq:Etrsym} is also applicable for other methods that use $JKL$-only integrals such as the antisymmetrized product of strongly orthogonal geminals, \cite{pernal:13ctc} the $\Delta$NO method, \cite{hollett2016cumulant,elayan2022deltano} and the recently proposed methodology based on Richardson-Gaudin states. \cite{fecteau2020reduced,fecteau2021richardson,johnson2023richardson} Another major advantage of including time-reversal symmetry is that the Piris-Ugalde orbital optimization algorithm can be applied to problems involving complex orbitals. This is because the diagonal terms of the gradient responsible for changing the orbital phase vanish ($g_{pp}=0$). This advantage can be further exploited in future implementations of the Piris RDMFT functional approximations and pCCD methods for extended systems where several complex one-body Bloch states are required. In such cases, quadratic convergent algorithms that require the Hessian matrix become computationally prohibitive. Therefore, this work sheds light on the technical and theoretical aspects encountered when implementing the Piris RDMFT functional approximations and pCCD methods for extended systems. \rev{On the other hand, restricting ourselves to $JK$-only functionals implies that the usual shortcomings of using two-index quantities to approximate the electron-electron interactions are still present. For instance, the well-known problem of the inability of these RDMFT functionals to account for long-range dynamic electronic correlation effects \cite{cioslowski2002density} (responsible for the so-called weak interactions) is not improved by using complex-valued orbitals and time-reversal symmetry. Future works are still needed to focus on these open issues. In particular, cost-effective approaches would be desirable as the currently available proposals based on perturbation theory \cite{piris2018dynamic,rodriguez2021coupling,pernal2014accurate,pernal2018electron,pastorczak2018correlation} can not be directly employed in large systems (e.g.~extended systems).}

From a practical perspective, our numerical examples reveal that the real and complex solutions may differ in regions where non-dynamic correlation effects are enhanced with complex energies lying below the real ones in such cases. In the case of HF calculations, the real solutions correspond to saddle points of the complex optimization problem. On the other hand, for RDMFT functional approximations and the pCCD method, the real solutions are local minima. Interestingly, the TGNOF energies may lie below the FCI ones in regions dominated by non-dynamic correlation effects, suggesting that this functional approximation introduces $N$-representability violations. \rev{All RDMFT functionals presented in this work use two-index quantities to approximate the 2RDM elements. Hence, ideally, the ``best'' functional approximation one can build should reproduce the doubly occupied configuration interaction (DOCI) results. \cite{weinhold1967reduced,weinhold:67jcp} However, the $N$-representability violations allow the TGNOF functional to produce total energies that lie below the DOCI (or the FCI) ones, especially in regions where non-dynamic electronic correlation effects are important. This unpleasant feature can deteriorate the performance of the TGNOF approximation as it may produce lower energy at the expense of increasing the $N$-representability violations. Nevertheless, we have observed that this effect is system-dependent as the \ce{BeH2} energies lie below the FCI ones, while the \ce{N2} energies remain far from the FCI values. For this reason, $N$-representability violations can limit the reliability of the TGNOF functional in practical applications, but these violations might not be decisive in all cases. However, future developments of RDMFT approximations should focus on overcoming this issue before using these approximations in extended systems.} Finally, we have shown that the THF solution corresponds to the non-relativistic limit of the KR-4c-DHF method, where the RHF solution is attained only when it is equivalent to the THF one in regions where the non-dynamic electronic correlation effects are not dominant.   

\begin{acknowledgements}
M.R.-M.~thanks the European Commission for a Horizon 2020 Marie Skłodowska-Curie Individual Fellowship (891647-ReReDMFT).
P.-F.L.~has received financial support from the European Research Council (ERC) under the European Union's Horizon 2020 research and innovation program (Grant agreement no.~863481).
\end{acknowledgements}

\section*{Supplementary Material}
The Supplementary Material for the present article includes the \ce{H2} case, where the RPNOF5 and TPNOF5 solutions coincide, \rev{as well as the pCCD results for the \ce{BeH2}.}

\section*{Data Availability}
The data that support the findings of this study are available within the article and its supplementary material.

\appendix  
\section{Appendix A: The real RDMFT and pCCD Hessian}
\label{app:hessian}
For real orbitals, the one- and two-electron integrals, the occupation numbers, second-order reduced density matrix elements, and the parameters $\kappa_{pq}$ are real. For restricted calculations, only the elements $\kappa_{pq}$ with $p>q$ are unique. Hence, we have
\begin{equation}
    \hkappa= \sum_{p>q} \sum_{\sigma} \kappa_{pq} (\cre{c}{p\sigma}\anh{c}{q\sigma}-\cre{c}{q\sigma}\anh{c}{p\sigma}),
\end{equation}
only the $g_{pq}$ for $p>q$ elements of the gradient are needed, and only $p>q$ and $r>s$ terms of the Hessian are required. The Hessian is generally a Hermitian matrix (symmetric in the real case); thus, its construction only requires building the upper diagonal part and applying the symmetry conditions. In the case of RDMFT and pCCD calculations, the construction of the Hessian scales as $M^5$, which makes it reasonable in terms of computational cost when compared with more complex methods where the construction of the Hessian scales as $M^7$. The Hessian for real orbitals in terms of the auxiliary ${\bf \tG}$ matrix elements takes the following form 
\begin{align}
G_{pq,rs}&=\pdv[2]{E(\bkappa)}{\kappa_{pq}}{\kappa_{rs}}= \tG_{pq,rs}- \tG_{qp,rs}-\tG_{pq,sr}+\tG_{qp,sr} 
\end{align}
that are evaluated using Eq.~\eqref{eq:Gpqrs_tilde}. Notice that the real Hessian is given by the first four elements of Eq.~\eqref{eq:hessian} because they correspond to partial derivatives taken with respect to $\Re \kappa_{pq}$ elements. Finally, for completeness, let us mention that the 6th to the 8th elements in Eq.~\eqref{eq:hessian} are obtained with partial derivatives with respect to $\Im \kappa_{pq}$ while the third line is produced by crossed $\Re \kappa_{pq}$  and $\Im \kappa_{pq}$ derivatives.   

\bibliography{general}

\end{document}